\pdfoutput=1
\documentclass[12pt, leqno]{article}
\usepackage[toc,page]{appendix}
\usepackage{amsmath,setspace,multirow,lineno}
\usepackage[font=small,labelfont=bf]{caption}
\usepackage[top=1.2in, bottom=1.2in, left=1.2in, right=1.2in]{geometry}
\usepackage[round]{natbib}
\usepackage{color, soul, bbm}

\DeclareCaptionStyle{italic}[justification=centering]{labelfont={bf},textfont={it},labelsep=colon}
\usepackage{graphicx,psfrag,epsf,textcomp,epstopdf, amsthm}
\usepackage{enumerate, dsfont}
\usepackage{natbib}
\usepackage{url,xcolor}
\usepackage{booktabs, subfig, bm, paralist, mathpazo, tikz, longtable, microtype}
\usepackage[linesnumbered,ruled,vlined]{algorithm2e}
\newenvironment{psmallmatrix}
{\left(\begin{smallmatrix}}
{\end{smallmatrix}\right)}

\usepackage[pdftex, colorlinks=true]{hyperref}
\definecolor{darkblue}{rgb}{0,0,.6}
\hypersetup{citecolor=darkblue,linkcolor=darkblue,urlcolor=darkblue}
\definecolor{DarkRed}{rgb}{.7,0,.4}
\definecolor{ao(english)}{rgb}{0.0, 0.5, 0.0}

\usepackage{comment,orcidlink}
\DeclareMathAlphabet\mathbfcal{OMS}{cmsy}{b}{n}

\newcommand{\blind}{0}

\newcommand{\Y}{\mathcal{Y}}

\graphicspath{{plots/}}

\newsavebox\CBox
\def\textBF#1{\sbox\CBox{#1}\resizebox{\wd\CBox}{\ht\CBox}{\textbf{#1}}}

\newtheorem{@definition}{\sc Definition}[section]

\newcommand{\bb}{\boldsymbol}

\begin{document}

\def\spacingset#1{\renewcommand{\baselinestretch}{#1}\small\normalsize} \spacingset{1}

\if0\blind
{
\title{\bf Forecasting high-dimensional functional time series with dual-factor structures}} 
   \author{Chen Tang \orcidlink{0000-0002-0948-6073} \footnote{Postal address: Research School of Finance, Actuarial Studies and Statistics, Level 4, Building 26C, Australian National University, Kingsley Street, Acton, Canberra, ACT 2601, Australia; Email: chen.tang@anu.edu.au.}\\
Research School of Finance, Actuarial Studies and Statistics \\
Australian National University \\
\\	
Han Lin Shang \orcidlink{0000-0003-1769-6430} \\
Department of Actuarial Studies and Business Analytics \\
Macquarie University \\
\\
Yanrong Yang \orcidlink{0000-0002-3629-5803} \\
Research School of Finance, Actuarial Studies and Statistics \\
Australian National University \\
\\
Yang Yang \orcidlink{0000-0002-8323-1490} \\
 School of Information and Physical Sciences\\
 The University of Newcastle
}
\date{}
\maketitle
\fi

\if1\blind
{
\title{\bf Forecasting high-dimensional functional time series with dual-factor structures}
} \fi

\date{\vspace{-5ex}}
\maketitle
   
\begin{abstract}
We propose a dual-factor model for high-dimensional functional time series (HDFTS) that considers multiple populations. The HDFTS is first decomposed into a collection of functional time series (FTS) in a lower dimension and a group of population-specific basis functions. The system of basis functions describes cross-sectional heterogeneity, while the reduced-dimension FTS retains most of the information common to multiple populations. The low-dimensional FTS is further decomposed into a product of common functional loadings and a matrix-valued time series that contains the most temporal dynamics embedded in the original HDFTS. The proposed general-form dual-factor structure is connected to several commonly used functional factor models. We demonstrate the finite-sample performances of the proposed method in recovering cross-sectional basis functions and extracting common features using simulated HDFTS. An empirical study shows that the proposed model produces more accurate point and interval forecasts for subnational age-specific mortality rates in Japan. The financial benefits associated with the improved mortality forecasts are translated into a life annuity pricing scheme.
\\

\end{abstract}

\newpage
\spacingset{1.58}

\section{Introduction}\label{sec:1}

Increased human life expectancy brings longevity risks to governments, health care systems and life insurance industries. Government stakeholders and industry practitioners need accurate and easily understandable mortality forecasting methods to help them gauge the risks associated with extended longevity. Modelling and forecasting age-specific mortality rates have become an ongoing research endeavour, noting particularly the seminal work by \cite{lee1992modeling} advancing the field of research. However, the Lee-Carter model and many of its extensions and modifications \citep[see, e.g.,][]{renshaw2003lee,li2013extending,wisniowski2015bayesian} consider only a single population. A common disadvantage of such univariate mortality modelling methods is that they tend to produce divergent forecasts for a group of related populations over relatively long forecast horizons. As a remedy, coherent mortality forecasting methods \cite[see, e.g.,][]{pampel2005forecasting,  li2005coherent, li2013poisson} that simultaneously consider multiple populations have gained popularity in recent years. This is because jointly modelling multiple closely related mortality series facilitates the pooling of common information of the considered populations and hence results in improved forecasts \citep[][]{shang2016mortality, CSS24}.

However, the aforementioned multivariate mortality modelling methods cannot handle high-dimensional multi-population mortality data. When working with mortality data in which the number of populations (cross-sectional units) is comparable to, or even more significant than, the observation years (sample size of FTS), many conventional statistical tools tend to fail, and high-dimensional data analysis techniques are needed \citep[see, e.g.,][]{cai2010highdimensional,buhlmann2011statistics}. In addition, various populations (e.g., different countries) may report mortality rates over different time intervals. Limiting the mortality observations to the same time is often necessary before applying the conventional matrix-based multivariate mortality models \citep[e.g., the extended Lee-Carter model of][]{li2013extending}. 

To solve these problems, we propose to model multi-population mortality time series in a functional data analysis framework with high-dimensional data analysis techniques. In this functional data analysis setting, the age-specific mortality rates at any given year can be regarded as a curve; hence the curves relate to any particular population and are recorded sequentially according to time form FTS \citep[see, e.g.,][]{ramsay2006functional, hormann2010weakly}. In the analysis of FTS, dimension-reduction techniques, such as functional principal component analysis (FPCA), are necessary to collapse the infinite-dimensional curves into a series of orthonormal basis functions and their corresponding matrix-valued coefficients. To handle the high dimensionality of data related to a large number of populations, factor analysis is often used to reduce the variation contained in multi-population data into a smaller number of uncorrelated factors \citep[see, e.g.,][]{bai2008large}. When applying our proposed method to multi-population mortality data, we employ both dimension-reduction techniques to solve the two-fold ``curse of dimensionality" \citep{bellman1966dynamic}.

The key contributions of this study are as follows.
\begin{enumerate}
    \item[1)] We propose a dual-factor structure for modelling and forecasting high-dimensional functional time series (HDFTS). The proposed method effectively decomposes HDFTS into the product of a vector of common loading (front-loading) functions, a low-dimensional matrix-valued time series, and a finite collection of factor loading (back-loading) functions. When applied to the analysis of high-dimensional mortality data, our method retains common age characteristics of human mortality curves in the front-loadings and identifies region-specific features over age in the back factor loadings. In addition, temporal dynamics of the original HDFTS are captured by the low-dimensional matrix-valued time series that facilitates forecasting.
    \item[2)]  The proposed general-form dual-factor structure covers several commonly used functional factor models as special cases. By selecting the appropriate front and back-loading functions, the proposed general dual-factor model structure can be extended to several existing models in the literature. 
    \item[3)] Using a Monte Carlo simulation study, we demonstrate the finite-sample performance of the proposed method in recovering cross-sectional heterogeneous basis functions and extracting common features.
    \item[4)] In an empirical analysis of Japanese subnational mortality rates, we further demonstrate the HDFTS model's ability to produce more accurate point and interval forecasts than several commonly used multi-population mortality models. Moreover, we demonstrate the financial impact of the accurate forecasts produced by our model on life annuity pricing. It is shown that the proposed model could result in significant savings for the insurance providers' reserve funds. 
\end{enumerate}

\section{Related work}

Early attempts at jointly modelling multivariate FTS in the literature, include, for example, the work of \cite{di2009multilevel}, which introduced the multilevel FPCA that performs the conventional FPCA on intra- and inter-subject geometric components of multilevel functional data. \cite{shang2016mortality} applied multilevel FPCA to model the mortality rates of multiple populations. To consider functional data exhibiting significant heterogeneity, \cite{tang2020clustering} proposed a clustering method capable of grouping multiple FTS into homogeneous sets before jointly modelling the curves in each set.  \cite{gao2018high} introduced a two-step approach for reducing the HDFTS into a small number of latent factors with corresponding factor loadings and basis functions. To obtain a global FPCA projection for multi-population FTS, \cite{nisol2019high} arranged HDFTS into a panel structure and performed an eigendecomposition to the stacked FTS. \cite{ZD23} derived Gaussian and multiplier bootstrap approximations for sums of HDFTS. 

Parallel to the functional framework but still focusing on multi-population data, \cite{wang2019factor} proposed a factor model for matrix-valued high-dimensional time series. Viewing the matrix input across columns and rows reveals information on dependence from two distinct perspectives. In the context of multi-population mortality data, all the considered ages can be arranged as row variables, whereas labels of the specific populations can be used as column variables. The model uses the mortality observations in any particular year to fill in the cells of the corresponding matrix and repeats the process for each observation time to form a matrix-valued time series. The correlation of mortality rates in the age direction and among the cross-sectional units can then be summarized via factor analysis into low-dimensional front-loading and back-loading matrices. All temporal information of the original data is retained in a low-dimensional matrix-values time series in the middle of the two loading matrices. This method developed by \cite{wang2019factor} is conceptually similar to our dual-factor structure for HDFTS, introduced in Section~\ref{sec:2}, but considers conventional discrete-valued time series data. 

Our proposed model has several advantages over the existing methods.
\begin{asparaenum}
\item[1)] Our dual-factor model for HDFTS considers mortality observations as finite realizations of an underlying continuous process with age as a continuum. Under this functional framework, we can better capture mortality dependence across various ages for a particular population than can the matrix-valued time series factor model of \cite{wang2019factor}. 
\item[2)] Compared to existing models for HDFTS, our proposed two-fold dimension reduction algorithm can effectively isolate the homogeneous features for all data and the heterogeneous basis functions specific to each population, with the majority of the temporal dependence of the original HDFTS retained in a series of low-dimensional factor matrices. This separation of homogeneity and heterogeneity in HDFTS results in accurate forecasts for multi-population mortality data. On the one hand, the heterogeneous basis functions accommodate the unique characteristics of each population, reducing information loss in dimensionality reduction. On the other hand, the extracted homogenous features contribute to non-divergent forecasts for closely related populations. 
\item[3)] We employ the dynamic FPCA for dimensionality reduction to adequately capture the temporal dependence of HDFTS \citep[see, e.g.,][]{hormann2015dynamic,rice2017plug}. This ideal property of dynamic FPCA is the result of using a long-run covariance function to estimate functional loadings. To our knowledge, the dynamic FPCA is often the preferred FTS feature extraction method for forecasting purposes \citep{gao2018high, yang2022temporal}. 
\item[4)] Our proposed model is highly flexible such that it can be viewed as a general case of several existing models by \cite{di2009multilevel}, \cite{gao2018high} and \cite{nisol2019high}. By specifying appropriate front- and back-loading functions, our model can be reduced to these popular multi-population FTS models. 
\end{asparaenum}

\cite{chang2024} recently proposed a two-step procedure for modelling high-dimensional HDFTS (TSHDFTS) that initially performs independent component analysis of functional observations to achieve segmentations. Any two curves after transformation are uncorrelated, both cross-sectionally and temporally. The second step involves a latent decomposition to identify a finite-dimensional dynamic representation for each transformed curve. This modelling framework by \cite{chang2024} may be viewed as a reduced form of \eqref{eq:4} where the front-loading function is simplified to a constant matrix. Information loss is inevitable in this step of reducing functions to scalers. In contrast, our proposed dual-factor model can better capture the age-varying features of mortality rates via the front-loading functional factor. The empirical results in Section~\ref{sec:6} verify that our proposed method produces more accurate point and interval forecasts than TSHDFTS.

\cite{CSS24} considered a two-way functional ANOVA model for HDFTS that decomposes observations into a deterministic component capturing differences of the functional means and a time-varying residual component gauging the dynamic structure of the curves. Compared to the two-way functional ANOVA method, our proposed dual-factor model has a more flexible form, imposing fewer constraints on the functional means. Moreover, our dual-factor model facilitates a decomposition of time-varying functional observations into common factors and their associated back-loading functions. The back-loading functions improve the interpretability of our results, while the common factors summarize time trends in human mortality functions.

The rest of the paper is organized as follows. Section~\ref{sec:2} details our factor model and draws connections with several existing models. Section~\ref{sec:3} outlines our estimation procedures. Section~\ref{sec:4} describes our forecasting method based on the factor model for HDFTS. Section~\ref{sec:5} presents Monte-Carlo simulation studies to show the finite-sample performance of the proposed method on recovering the homogeneous and heterogeneous features of HDFTS. Section~\ref{sec:6} presents an empirical analysis of the Japanese subnational age-specific mortality data. Section~\ref{sec:7} concludes the paper, highlighting how the methodology can be extended.

\section{Factor model for HDFTS}\label{sec:2}

We propose a new dual-factor model for HDFTS, which separates the homogeneity and heterogeneity within HDFTS. The homogeneity represents the similarity that is shared among all cross sections, while the heterogeneity represents the dissimilarity. In doing so, we reduce the dimensions of the HDFTS, incuding summarizing the temporal information of the original data into a low-dimensional matrix-valued time series. Doing so addresses the curse of dimensionality and makes forecasting feasible.

\subsection{Functional panel data structure}

Let $\{\mathbfcal{X}_{t}: t = 1, \ldots, T\}$ be an $N$-dimensional FTS, where $N$ is the number of sets of FTS, and $T$ is the sample size or the number of functions within each set of FTS. The HDFTS can be organized into a single matrix format such that 
\begin{equation*}
\mathbfcal{X}(u) = 
  \begin{bmatrix}
  \mathcal{X}_1^{(1)}(u) & \mathcal{X}_1^{(2)}(u) &\ldots & \mathcal{X}_1^{(N)}(u)\\
  \mathcal{X}_2^{(1)}(u) & \mathcal{X}_2^{(2)}(u) & \ldots & \mathcal{X}_2^{(N)}(u)\\
  \vdots & \vdots & \ddots & \vdots\\
  \mathcal{X}_{T-1}^{(1)}(u) & \mathcal{X}_{T-1}^{(2)}(u) & \ldots & \mathcal{X}_{T-1}^{(N)}(u)\\
  \mathcal{X}_T^{(1)}(u) & \mathcal{X}_T^{(2)}(u)& \ldots & \mathcal{X}_T^{(N)}(u)\\
\end{bmatrix},
\end{equation*}
where $\mathbfcal{X}(u)$ is a $T\times N$ matrix, each entry in the matrix, $\mathcal{X}_t^{(i)}$ for $i\in\{1, \ldots, N\}$, is defined in $\bb{\mathcal{H}}:= \mathcal L^{2}(\mathcal{I})$, a Hilbert space of square-integrable functions on a real interval $\mathcal{I} \in [a,b]$, with the inner product  $\langle f, g\rangle = \int_{\mathcal{I}} f(u)g(u)$ and the norm $\|\cdot\| = \langle \cdot, \cdot\rangle ^{\frac{1}{2}}$. 

Under our setting, the $i^{\textsuperscript{th}}$ column of $\mathbfcal{X}(u)$, $\mathbfcal{X}^{(i)}(u) = [\mathcal{X}_1^{(i)}(u), \ldots, \mathcal{X}_T^{(i)}(u)]^{\top}$, represents a FTS, which contains serial dependence. The $t^{\textsuperscript{th}}$ row of $\mathbfcal{X}(u)$, $\mathbfcal{X}_{t}(u) = [\mathcal{X}_t^{(1)}(u), \ldots, \mathcal{X}_t^{(N)}(u)]$, consists of functions of all cross sections $N$ at a specific time $t$. It is noteworthy that these are correlated. Let us consider the age-specific mortality rate as an example, each entry in $\mathbfcal{X}(u)$ represents the age-specific mortality rate curve of a specific region at a specific time. The $i^{\textsuperscript{th}}$ column of $\mathbfcal{X}(u)$ represents the mortality rate curves of region $i$ across the sample period, and the $t^{\textsuperscript{th}}$ row of $\mathbfcal{X}(u)$ represents the mortality rate curves of all countries at time $t$.

\subsection{Model setup}

Denote $\mathcal{Y}_{t}^{(i)}(u) = \mathcal{X}_{t}^{(i)}(u) - \mu^{(i)}(u)$ as the centered FTS. Let us consider the following factor model for the centred HDFTS 
\begin{equation}\label{eq:1}
  \mathcal{Y}_{t}^{(i)}(u) =  \bb{\phi}(u)\bb{F}_t [\bb{\lambda}^{(i)}(u)]^{\top} + \epsilon_{t}^{(i)}(u), 
\end{equation}
where $\bb{\phi}(u)$ is a vector of length $r$, with each entry being function-valued, that is, the features common to all sets of FTS; $\bb{F}_t$ is an $r\times k$ matrix; $\bb{\lambda}^{(i)}(u)$ is a vector of length $k$, with each entry being function-valued, that is, specific for each $i\in\{1, 2, \ldots, N\}$, which reflects the specific characteristic (heterogeneity) of each FTS; and $\epsilon_{t}^{(i)}(u)$ is a white noise idiosyncratic process that has no serial dependence \citep[see, e.g.,][]{lam2011estimation,lam2012factor}. We term $\bb{\phi}(u)$ and $\bb{\lambda}^{(i)}(u)$ as front-loading and back-loading, respectively \citep[see also][]{wang2019factor}. The difference is that the factor loadings are scalar-valued in their work, while in our proposed model the factor loadings are function-valued. This extension allows our model to be more general. 

\subsection{Model interpretation}

The proposed model can be interpreted in the following ways. We use the back-loadings to characterize the heterogeneity of the cross sections. Separating the heterogeneity reduces the original data into a multivariate FTS of smaller dimensions. These common multivariate FTS characterize the homogeneity of the original data. After reducing the dimensions of these common multivariate FTS, the temporal information is captured by low-dimensional matrix-valued time series, which facilitates forecasting. 

\subsubsection{Dimension reduction along cross sections}

Denoting $ \bb{\mathcal{F}}^{\top}_t(u) = \bb{\phi}(u)\bb{F}_t$, then the model in ~\eqref{eq:1} can be expressed as
\begin{equation}\label{eq:2} 
   \mathcal{Y}_{t}^{(i)}(u) = \bb{\lambda}^{(i)}(u)\bb{\mathcal{F}}_t(u) + \epsilon_{t}^{(i)}(u),
 \end{equation} 
where $\bb{\lambda}^{(i)}(u)$ is a $k$-dimensional functional factor loading vector, $\bb{\lambda}^{(i)}(u) = [\lambda_1^{(i)}(u), \ldots, \lambda_k^{(i)}(u)]$, with each entry being a function, defined previously, and $\bb{\mathcal{F}}_t(u)$ is a $k\times 1$ vector of functional factors at time $t$. This model differs from that of \cite{nisol2019high}, where the factor loading is function-valued, and the factors are scalar. The front-loading $\bb{\lambda}^{(i)}(u)$ is dependent on $i$, which represents the heterogeneity of each FTS, while the functional factor $\bb{\mathcal{F}}_t(u)$ is common to all $i$, which represents the common features of the HDFTS. Then, $\bb{\mathcal{F}}(u) = [\bb{\mathcal{F}}_1(u), \ldots, \bb{\mathcal{F}}_T(u)]^{\top}$ is a $k$-dimensional FTS ($k\ll N$). By doing this, we separate the common features of the HDFTS, denoted by $\bb{\mathcal{F}}(u)$, from the heterogeneity, denoted by $\bb{\lambda}^{(i)}(u)$, and reduce the HDFTS to FTS with fewer dimensions.

\cite{bai2009panel} proposed a panel data model with interactive fixed effects, where the multiplication of the $r$-dimensional factor back-loading $\bb{\lambda}_i$ and common factors $\bb{F}_t$ can be viewed as the interaction of these two terms. The interactive effects model is more general than the additive effects model. It allows the individual effect of $i$ and the time effect of $t$ to be involved in the model interactively rather than additively, giving more flexibility. \cite{freyberger2018non} argued that the interactive fixed effects model could handle cases where heterogeneity is not time-homogeneous. \cite{tang2020clustering} proposed a functional panel data model with additive fixed effects. To handle time-varying heterogeneity, they further proposed a clustering algorithm to search for homogeneous subgroups before jointly modelling the FTS. Our model can be viewed as a functional extension of the interactive fixed-effects model, as stated in~\eqref{eq:2}, where both the factor loading $\bb{\lambda}^{(i)}(u)$ and the common factors $\bb{\mathcal F}_t(u)$ are function-valued, and $\bb{\lambda}^{(i)}(u)$ captures the specific effects of different cross sections and $\bb{\mathcal F}_t(u)$ denotes the common features with the multiplication of $\bb{\lambda}^{(i)}(u)$ and $\bb{\mathcal F}_t(u)$ allowing for time-varying heterogeneity.
  
\subsubsection{Dimension reduction on infinite-dimensional functions}

After applying dimension reduction, the FTS $\bb{\mathcal{F}}^{\top}_t(u)$ are obtained. Then, a functional factor model can be applied such that 
\begin{equation}\label{eq:3}
  \bb{\mathcal{F}}^{\top}_t(u) = \bb{\phi}(u)\bb{F}_t + \bb{\upsilon}_{t}(u),
\end{equation}
where $\bb{\phi}(u)$, the common functional loading, is a vector of length $r$; $\bb{F}_t$ is an $r\times k$ matrix and $\bb{\upsilon}_{t}(u)$ is also the white noise idiosyncratic component. In doing so, all the temporal dynamics are reflected in the matrix-valued time series  $\{\bb{F}_t\}$.

Combining~\eqref{eq:2} and ~\eqref{eq:3}, we can obtain the proposed model as in~\eqref{eq:1}. The HDFTS $\mathbfcal{Y}_{t}(u) = [\mathcal{Y}_t^{1}(u), \ldots, \mathcal{Y}_t^{N}(u)]^{\top}$ can also be expressed in a matrix format, 
\begin{equation}\label{eq:4}
  \mathbfcal{Y}_{t}(u) = \bb{\phi}(u)\bb{F}_t [\bb{\Lambda}(u)]^{\top} + \bb{\epsilon}_{t}(u),
\end{equation}
where the systems of back-loadings $\bb{\Lambda}(u)=[\Lambda^{(1)}(u),\dots,\Lambda^{(N)}(u)]$ are an ($N\times k, \, k\ll N$) matrix, with the $i^{\text{th}}$ row being $\bb{\lambda}^{(i)}(u)$; and $\bb{\epsilon}_{t}(u)$ is an $N$-dimensional vector of the white noise processes. 

To better understand the model, we can extend the first component on the right-hand side of~\eqref{eq:4} as
{\footnotesize{ 
\begin{equation}\label{eq:5}
\begin{split} 
  \bb{\phi}(u)\bb{F}_t [\bb{\Lambda}(u)]^{\top} = & \color{ao(english)}{\begin{bmatrix}
    \phi_1(u) & \phi_2(u) & \ldots & \phi_r(u)
  \end{bmatrix}}\color{red}{\begin{bmatrix}
    F_{t,11} & F_{t,12} & \ldots & F_{t,1k}\\
    F_{t,21} & F_{t,22} & \ldots & F_{t,2k}\\
    \vdots & \vdots & \ddots & \vdots \\
    F_{t,r1} & F_{t,r2} & \ldots & F_{t,rk}\\
  \end{bmatrix}}\color{blue}{\begin{bmatrix}
    \lambda_{1}^{(1)}(u) & \lambda_{1}^{(2)}(u) &\ldots & \lambda_{1}^{(N)}(u)\\
    \lambda_{2}^{(1)}(u)& \lambda_{2}^{(2)}(u) &\ldots & \lambda_{2}^{(N)}(u)\\
    \vdots & \vdots & \ddots & \vdots \\
    \lambda_{k}^{(1)}(u) & \lambda_{k}^{(2)}(u) &\ldots & \lambda_{k}^{(N)}(u)\\
  \end{bmatrix}}\\
  = & \color{blue}{\begin{bmatrix}
    \displaystyle \sum_{p=1}^{r}\sum_{q=1}^{k}\phi_p(u)F_{t,pq} \lambda_q^{(1)}(u) & \displaystyle \sum_{p=1}^{r}\sum_{q=1}^{k}\phi_p(u) F_{t,pq} \lambda_q^{(2)}(u) & \ldots & \displaystyle \sum_{p=1}^{r}\sum_{q=1}^{k}\phi_p(u)F_{t,pq}\lambda_q^{(N)}(u)  
  \end{bmatrix}}
\end{split}
\end{equation}
}}
\hspace{-.1in} where $\bb{\phi}(u)$ is the loading functions common to all cross sections. $\bb{\Lambda}(u)$ is the system loading curves specific to different cross sections. By reducing the dimensions of both functional continuum and cross sections, we can extract the important information of the HDFTS into an $r\times k\times T$ factor array $\bb{F}$. This array forms a matrix-valued time series, $\{\bb{F}_t, t = 1, 2, \ldots, T\}$, which contains all the temporal dynamics of the HDFTS.

\subsubsection{Connections with other existing models}

The generality of the proposed model lies in the flexibility of the choice of the front- and back-loading functions. The connections with other existing models are discussed here.

\textbf{Case 1 (Additive Fixed-Effect Model, \cite{di2009multilevel})}: Setting each entry in the back-loading being the same constant function, that is, $\lambda^{(i)}_{q}(u) = c$ for all $q \in 1, 2, \ldots, k$ and $i \in 1, 2, \ldots, N$. Equation~\eqref{eq:5} can be reduced as 
\begin{equation*}
\begin{split} 
  \bb{\phi}(u)\bb{F}_t \bb{\Lambda}^{\top}(u) 
  & = \begin{bmatrix}
    \displaystyle \sum_{p=1}^{r}\phi_p(u)\sum_{q=1}^{k}cF_{t,pq}  & & \displaystyle \sum_{p=1}^{r}\phi_p(u) \sum_{q=1}^{k}cF_{t,pq}  & & \ldots & & \displaystyle \sum_{p=1}^{r}\phi_p(u)\sum_{q=1}^{k}cF_{t,pq}
  \end{bmatrix}
  \end{split}
\end{equation*}
Let $\beta_{t,p} = \sum_{q=1}^{k}cF_{t,pq}$, then $\mathcal{X}_{t}^{(i)}(u)$ can be expressed as 
\begin{equation*}
  \mathcal{X}_{t}^{(i)}(u) = \mu^{(i)}(u)+ \sum_{p=1}^{r}\beta_{t,p}\phi_p(u) + \text{E}_{t}^{(i)}(u), 
\end{equation*}
where $\text{E}_{t}^{(i)}(u)$ is the error component, for ease of notation, let $\text{E}_{t}^{(i)}(u)$ represent the error component for different models described below. This is the special case of a functional panel data model with additive fixed effects as in \cite{di2009multilevel}, where there is no component for the cross section, $i$.

\textbf{Case 2 (Multivariate FPCA)}: If we set each entry in the back-loading matrix as a constant function specific to each $i$, i.e., $\lambda^{(i)}_{q}(u) = c_i$,  for all $q \in 1, 2, \ldots, k$. Equation~\eqref{eq:5} can be reduced as 
\begin{equation*}
\begin{split} 
  \bb{\phi}(u)\bb{F}_t \bb{\Lambda}^{\top}(u) 
  & = \begin{bmatrix}
    \displaystyle \sum_{p=1}^{r}\phi_p(u)\sum_{q=1}^{k}c_1F_{t,pq}  & & \displaystyle \sum_{p=1}^{r}\phi_p(u) \sum_{q=1}^{k}c_2F_{t,pq}  & & \ldots & &\displaystyle \sum_{p=1}^{r}\phi_p(u)\sum_{q=1}^{k}c_NF_{t,pq}
  \end{bmatrix}
  \end{split}
\end{equation*}
Let $\beta_{t,p}^{(i)} = \sum_{q=1}^{k}c_iF_{t,pq}$, then $\mathcal{X}_{t}^{(i)}(u)$ can be expressed as 
\begin{equation*}
  \mathcal{X}_{t}^{(i)}(u) = \mu^{(i)}(u)+ \sum_{p=1}^{r}\beta_{t,p}^{(i)}\phi_p(u) + \text{E}_{t}^{(i)}(u), \quad t = 1,2,\ldots, T,
\end{equation*}
then the model becomes the Karhunen-Lo\`{e}ve representation for multivariate functional data, where all functions from different cross sections share the same bases.

\textbf{Case 3 (High-Dimensional Factor Model, \cite{nisol2019high})}: If we set each entry in the back-loading matrix as a function specific to $i$, that is, $\lambda^{(i)}_{q}(u) = \gamma^{(i)}(u)$,  for all $q \in 1, 2, \ldots, k$. Equation~\eqref{eq:5} can be reduced as 
{\small{
\begin{equation*}
\begin{split} 
  \bb{\phi}(u)\bb{F}_t \bb{\Lambda}^{\top}(u) 
  & = \begin{bmatrix}
    \displaystyle \sum_{p=1}^{r}\gamma^{(1)}(u)\phi_p(u)\sum_{q=1}^{k}F_{t,pq} & & \displaystyle \sum_{p=1}^{r}\gamma^{(2)}(u)\phi_p(u) \sum_{q=1}^{k}F_{t,pq}  & &\ldots & &\displaystyle \sum_{p=1}^{r}\gamma^{(N)}(u)\phi_p(u)\sum_{q=1}^{k}F_{t,pq}
  \end{bmatrix}
  \end{split}
\end{equation*}
}}
Let $\beta_{t,p} = \sum_{q=1}^{k}F_{t,pq}$ and $\phi_p^{(i)}(u) = \gamma^{(i)}(u)\phi_p(u)$, 
$\mathcal{X}_{t}^{(i)}(u)$ can be expressed as 
\begin{equation*}
  \mathcal{X}_{t}^{(i)}(u) = \mu^{(i)}(u)+ \sum_{p=1}^{r}\beta_{t,p}\phi_p^{(i)}(u)+ E_{t}^{(i)}(u), \quad t = 1,2,\ldots, T,
\end{equation*}
then it is the high-dimensional factor model of \cite{nisol2019high}. Similarly, we can obtain the same model if the front-loading is set to be a constant scalar, i.e., $\phi_p(u) = c$. 

\textbf{Case 4 (Individual FPCA, \cite{gao2018high})}: If we set each entry in the front-loading being a constant specific to $i$ for $i = 1, 2, \ldots, N$, that is, $\phi_p(u) = c_i$ for all $p = 1, 2, \ldots, r$. Equation~\eqref{eq:5} can be re-expressed as 
\begin{equation*}
\begin{split} 
  \bb{\phi}(u)\bb{F}_t \bb{\Lambda}^{\top}(u) 
  & = \begin{bmatrix}
    \displaystyle \sum_{q=1}^{k}\lambda_q^{(1)}(u)\sum_{p=1}^{r}c_1F_{t,pq}  & & \displaystyle \sum_{q=1}^{k}\lambda_q^{(2)}(u)\sum_{p=1}^{r}c_2 F_{t,pq}  & & \ldots & &\displaystyle \sum_{q=1}^{k}\lambda_q^{(N)}(u)\sum_{p=1}^{r}c_NF_{t,pq}
      \end{bmatrix}
  \end{split}
\end{equation*}
Let $\beta_{t,q}^{(i)} = \sum_{p=1}^{r}c_i F_{t,pq}$ and 
$\phi_q^{(i)}(u) = \lambda_q^{(i)}(u)$, $\mathcal{X}_{t}^{(i)}(u)$ can be expressed as 
\begin{equation*}
  \mathcal{X}_{t}^{(i)}(u) = \mu^{(i)}(u)+ \sum_{q=1}^{k}\beta_{t,q}^{(i)}\phi_q^{(i)}(u)+ E_{t}^{(i)}(u), \quad t = 1,2,\ldots, T,
\end{equation*}
then it is the Karhunen-Lo\`{e}ve representation for univariate functional data, from which the factor model of \cite{gao2018high} is derived.

\textbf{Case 5 (Matrix Factor Model, \cite{wang2019factor})}: If we discretize the functional continuum~$u$ into grid points and consider the discrete high-dimensional setting rather than the functional setting, and set each entry in the back-loading matrix is a different scalar, that is, $\lambda_q^{(i)}(u) = c_{i,q}$, then it is equivalent to the factor model for high-dimensional matrix-valued time series of \cite{wang2019factor}.

\section{Parameter estimation}\label{sec:3}

We present an estimation procedure for parameters $\{\mu^{(i)}(u), \bb{\phi}(u), \bb{F}_t\}$ in~\eqref{eq:1}. In general, the idea of the estimation procedure is to estimate the systems of back-loading first so that the HDFTS is reduced to the FTS of lower dimensions. We then proceed to estimate the common front-loading and the matrix-valued time series, $\bb{F_t}$. 

Suppose we have $T$ curves in each cross section, the estimation of $\mu^{(i)}(u)$ is as follows
\begin{equation*}
\widehat\mu^{(i)} (u)= \frac{1}{T}\sum_{t=1}^{T}\mathcal{X}_{t}^{(i)}(u).
\end{equation*}
Then, $\mathcal{Y}_{t}^{(i)}(u)$ is expressed as
\begin{equation*}
\mathcal{Y}_{t}^{(i)}(u) = \mathcal{X}_{t}^{(i)}(u) - \widehat\mu^{(i)} (u).
\end{equation*}

\subsection{Estimation of functional loadings \texorpdfstring{$\bb{\Lambda(u)}$}{}}

The cross-sectional specific functional back-loadings $\{\bb{\lambda^{(i)}(u)}\}$ represent the heterogeneity of each FTS. To accommodate the serial dependence in each cross section, $\bb{\lambda^{(i)}(u)}$ is estimated by the eigendecomposition of the long-run covariance operator corresponding to the $i^{\textsuperscript{th}}$ cross section.

Given the inadequacy of classic FPCA in accounting for the essential information provided by the serial dependence structure in FTS, dynamic FPCA is developed. \cite{horvath2013estimation} and \cite{panaretos2013fourier} defined smoothed periodogram-type estimates of the long-run covariance and spectral density operators for FTS. \cite{hormann2015dynamic} used the spectral density operator to create functional filters to construct mutually uncorrelated dynamic FPCs so that the dynamic FPC scores can be analyzed component-wise. \cite{rice2017plug} proposed a bandwidth selection method for estimates of the long-run covariance function based on finite-order weight functions that aim to minimize the estimator's asymptotic mean-squared normed error. Following their work, the long-run covariance function for the $i^{th}$ cross section $c^{(i)}(u,v)$ can be estimated by
\begin{equation*}
  \widehat c^{(i)}(u,v) = \sum_{s=-\infty}^{\infty}W\Big(\frac{s}{h}\Big) \widehat c^{(i)}_s(u, v),
\end{equation*}
where
\begin{align*}
\widehat c^{(i)}_s(u, v) = \left\{ \begin{array}{cc} 
                \frac{1}{T-s}\sum_{t=1}^{T-s}\widehat\mathcal{Y}_{t}^{(i)}(u)\widehat\mathcal{Y}^{(i)}_{t+s}(v), & s\geq 0;\\
                \frac{1}{T-s}\sum_{t=1-s}^{T}\widehat\mathcal{Y}_{t}^{(i)}(u)\widehat\mathcal{Y}^{(i)}_{t+s}(v),&  s< 0.\\
                \end{array} \right.
\end{align*}
$W(\cdot)$ is the kernel function that assigns different weights to the auto-covariance functions of different lags, and $h$ is the bandwidth. Despite many different types of kernel functions \cite[see][for types of kernel functions]{hansen1982large,white2014asymptotic,newey1986simple,andrews1991heteroskedasticity,gallant2009nonlinear}, it is common to assign more weight to the auto-covariance functions of small lags and less weight to the auto-covariance functions of large lags. Here, we use flat-top kernel functions as they give a reduced bias and faster rates of convergence \citep{politis1996flat,politis1999multivariate}. Flat-top kernels are of the following form
\begin{align*}
  W\Big(\frac{s}{h}\Big) = \left\{ \begin{array}{lll}
                1, & 0\leq|\frac{s}{h}|< m; \\ 
                 \frac{|\frac{s}{h}|-1}{m-1}, & m\leq|\frac{s}{h}|< 1;\\
                0, &  |\frac{s}{h}|\geq1,\\
                \end{array} \right.
\end{align*}
where $m<1$. The choice of bandwidth can greatly affect the finite-sample performance. Therefore, we apply the adaptive bandwidth selection procedure of \cite{rice2017plug} to gain a better estimate of the long-run covariance functions.

The long-run covariance operator can be estimated by
\begin{equation*}
  \widehat C^{(i)}(y)(u) = \sum_{s=-\infty}^{\infty}W\Big(\frac{s}{h}\Big) \widehat C^{(i)}_s(y)(u),
\end{equation*}
where $\widehat C^{(i)}_s(y)(u) = \int_{\mathcal{I}}\widehat c^{(i)}_s(u, v)y(v)dv$, is the sample covariance operator at lag $s$. 

The estimates of $\bb{\lambda^{(i)}(u)}$, denoted by $\bb{\widehat \lambda^{(i)}(u)} = [\widehat \lambda^{(i)}_1(u), \ldots, \widehat \lambda^{(i)}_{k_i}(u)]$, are the eigenfunctions corresponding to the first $k_i$ largest eigenvalues of $\widehat C^{(i)}(y)(u)$. Since we allow the $\bb{\lambda^{(i)}(u)}$ to vary according to $i$, it represents the heterogeneity of each population.

To maintain the matrix format of the loadings of each population, $k$ is selected to be 
\begin{equation*}
   k = {\operatorname{max}}\Big\{k_i:1\leq i\leq N\Big\},
\end{equation*}
where $k_i$ is the retained number of components used for $i^{th}$ population. 

The selection of the optimal number of functional principal components has been studied widely in the literature \citep{rice1991estimating,yao2005functional,hall2006assessing,chiou2012dynamical,hormann2015note}. In this paper, $k_i$ is determined by the cumulative percentage of variance method \citep{chiou2012dynamical}, such that
\begin{equation}\label{eq:6}
  k_i = \underset{k_i : k_i \geq 1}{\operatorname{argmin}}\left\{\frac{\sum_{m=1}^{k_i}\widehat\lambda^{(i)}_m}{\sum_{m=1}^{\infty}\widehat\lambda^{(i)}_m \mathbbm{1}\{\widehat\lambda^{(i)}_m>0\}}\geq P\right\},
\end{equation}
where $\lambda^{(i)}_m$ is the $m^{th}$ largest eigenvalue of $\widehat C^{(i)}(y)(u)$, $\mathbbm{1}\{\cdot\}$ denotes the binary indicator function and $P$ is a pre-determined value and is chosen to be 0.9 \citep[see, e.g.,][]{horvath2012inference}. 

\subsection{Estimation of \texorpdfstring{$\bb{\mathcal{F}}_t(u)$}{}}

In order to estimate $\bb{\phi(u)}$ and $\bb{F}_t$, we need to estimate $\bb{\mathcal{F}}_t(u)$ in~\eqref{eq:2}, that is, extract the common features of the HDFTS. Expressing~\eqref{eq:2} in matrix format, 
\begin{equation*}
\mathbfcal{Y}_{t}(u) = \bb{\Lambda}(u)\bb{\mathcal{F}}_t(u) + \bb{\epsilon}_{t}(u),
\end{equation*}
where $\bb{\Lambda}(u)$ is an $N\times k$ matrix of functions and $\bb{\mathcal{F}}_t(u)$ is a $k\times 1$ vector of functions. The $i^{\text{th}}$ entry of $\mathbfcal{Y}_{t}(u)$ can be expressed as follows:
\begin{equation*}
  \mathcal{Y}^{(i)}_t(u) = \sum_{q=1}^{k}\lambda_q^{(i)}(u)\mathcal{F}_{t,q}(u) + \epsilon_t^{(i)}(u).
\end{equation*}
This is essentially the functional concurrent regression model in \cite{james2009functional}, but without intercept function, where $\lambda_q^{(i)}(u)$ is a $k$-dimensional functional observation and $\mathcal{F}_{t,q}(u)$ is the functional regression coefficients. 

Let the $N$-dimensional residual functions vector be  
\begin{equation*}
   \bb{r_t(u)} = \bb{\mathcal{Y}_{t}(u)} - \bb{\Lambda}(u)\bb{\mathcal{F}}_t(u),
\end{equation*} 
then $\bb{\mathcal{F}}_t(u)$ can be estimated by minimizing the penalized sum of squares, such that
\begin{equation}\label{eq:7}
  \bb{\widehat{\mathcal{F}}_t(u)} = \underset{\bb{\mathcal{F}}_t(u)}{\operatorname{argmin}}\Big(\int_{\mathcal{I}}\bb{r_t^{\top}(u)}\bb{r_t(u)}du + \sum_{q=1}^{k}\gamma_q\int_{\mathcal{I}}[L_q\mathcal{F}_{t,q}(u)]^2du\Big),
\end{equation}
where $\gamma_q$ is the smoothing parameter and $L_q\mathcal{F}_{t,q}(u)$ is the linear differential operator applied to $\mathcal{F}_{t,q}(u)$. Together, $\gamma_q$ and $L_q\mathcal{F}_{t,q}(u)$ define the roughness penalty. 

Functional concurrent regression is performed for each $t = 1, 2, \ldots, T$, such that $\bb{\widehat\mathcal{F}}(u) = [\bb{\widehat\mathcal{F}}_1(u)^{\top}, \bb{\widehat\mathcal{F}}_2(u)^{\top}, \ldots, \bb{\widehat\mathcal{F}}_T(u)^{\top}]^{\top}$ is a matrix of function-valued objects, each with a dimension of $T\times k$, so that we have a panel of dimension-reduced FTS, which represents the common features.  

\subsection{Estimation of the front-loading function \texorpdfstring{$\bb{\phi(u)}$}{} and factor matrix \texorpdfstring{$\bb{F}_t$}{}}

Once $\bb{\widehat\mathcal{F}}(u)$ is obtained, we can estimate $\bb{\phi(u)}$ and $\bb{F}_t$ based on~\eqref{eq:3}. Since $\bb{\mathcal{F}}(u)$ is a dimension-reduced panel of FTS, where each column may be correlated, the key is to calculate the auto-cross-covariance of $\bb{\mathcal{F}}(u)$. \cite{wang2019factor} used the auto-cross-covariance to capture the co-movement observations from different cross sections of different lags. We extend it to the functional setting. Let $C_{h,lj}(u, v)$ be the auto-cross-covariance function of the $l^\text{th}$ and $j^\text{th}$ column of $\bb{\mathcal{F}}(u)$, for $l$ and $j = 1, \ldots, k$, such that 
\begin{equation*}
C_{h,lj}(u, v) = \text{Cov}\Big[\mathcal{F}_{t, l}(u), \mathcal{F}_{t+h, j}(u)\Big].
\end{equation*}
Since the operator using $C_{h,lj}(u, v)$ as a kernel is not nonnegative, we define a nonnegative operator \citep[using a similar idea as][]{bathia2010identifying},
\begin{equation*}
 M_{h,lj}(u,v) = \int_{\mathcal{I}}C_{h,lj}(u, z)C_{h,lj}(v, z)dz.
\end{equation*}
For a pre-determined integer $h_0$, the operator 
\begin{equation}\label{eq:8}
  M(u,v) = \sum_{h=1}^{h_0}\sum_{l=1}^{k}\sum_{j=1}^{k}M_{h,lj}(u,v),
\end{equation}
is also nonnegative. The auto-cross-covariance function can be estimated by
\begin{equation*}
  \widehat{C}_{h,lj}(u, v) = \frac{1}{T-h}\sum_{t=1}^{T-h}\widehat{\mathcal{F}}_{t, l}(u)\widehat{\mathcal{F}}_{t+h, j}(v).
\end{equation*}
Hence, the operator $M(u,v)$ can be estimated by
\begin{equation*}
  \widehat{M}(u,v) = \sum_{h=1}^{h_0}\sum_{l=1}^{k}\sum_{j=1}^{k}\int_{\mathcal{I}}\widehat{C}_{h,lj}(u, z)\widehat{C}_{h,lj}(v, z)dz.
\end{equation*}
The estimates of $\bb{\phi}(u)$, $\widehat{\bb{\phi}}(u) = [\widehat{\phi}_1(u), \ldots, \widehat{\phi}_r(u)]$, are the eigenfunctions corresponding to the first~$r$ largest eigenvalues of the eigen-decomposition on $\widehat{M}(u,v)$. The optimal $r$ can be selected using the same approach as in~\eqref{eq:6}. 

Once $\widehat{\bb{\phi}}(u)$ is obtained, the estimated $\widehat{\bb{F}_t}$ can be obtained through
\begin{equation*}
  \widehat{\bb{F}}_t = \int_{\mathcal{I}}\widehat{\bb{\phi}}(u)^{\top}\widehat{\bb{\mathcal{F}}}_t(u)^{\top}du,
\end{equation*}
where $\widehat{\bb{F}}_t$ is an $r\times k$ matrix. Therefore, all the temporal information is extracted into the $T$ sheets of $r\times k$ matrices.

In summary, we first separate the heterogeneity among the HDFTS, where a specific set of functional loading, $\widehat{\bb{\lambda}}^{(i)}(u)$, is obtained for each FTS. The common features can then be extracted into FTS of smaller dimensions, $\widehat{\bb{\mathcal{F}}}_t(u)$. Then, the functional dynamic factor model is applied to the dimension-reduced FTS, $\widehat{\bb{\mathcal{F}}}_t(u)$, such that all the temporal information can be extracted into the matrix-valued time series $\{\widehat{\bb{F}}_t: t = 1, \ldots, T\}$. 

\section{Forecasting}\label{sec:4}

Based on the proposed model, the functional panel data are extracted into $T$ sheets of fixed-dimensional factor matrices, $\bb{F}_t$. Forecasting the HDFTS is equivalent to forecasting the estimated factor matrices, $\widehat{\bb{F}}_t$. The $h$-step-ahead in sample forecast of the estimated factors, $\widehat{\bb{F}}_{\kappa+h|\kappa}$ can be obtained by fitting a vector autoregressive (VAR) model to the estimated factors $\{\widehat{\bb{F}}_{1}, \widehat{\bb{F}}_{2}, \ldots, \widehat{\bb{F}}_{\kappa}\}^{\top}$, where $\kappa <T$ is a sample size used for out-of-sample forecasting. In selecting the order of the VAR model, we follow the method of \cite{tsay2013multivariate}, where an information criterion, such as the Akaike information criterion, is used.

For a given sample size $\kappa$, the $h$-step-ahead in-sample forecast of the $i^\text{th}$ cross section in the HDFTS can be calculated as
 \begin{equation*}
   \widehat{\mathcal{X}}^{(i)}_{\kappa+h|\kappa}(u) = \widehat{\mu}^{(i)}(u) + \widehat{\bb{\phi}}(u)\widehat{\bb{F}}_{\kappa+h|\kappa}[\widehat{\bb{\lambda}}^{(i)}(u)]^{\top}.
 \end{equation*}
To capture the uncertainties associated with the point forecasts, point-wise prediction intervals  could also be constructed. We apply the nonparametric bootstrap approach of \cite{shang2018bootstrap} in constructing point-wise prediction intervals.

For a grid point $u_j$ on the curve, the $h$-step-ahead bootstrapped forecasts of the $i^\text{th}$ cross section, $\widehat{\mathcal{X}}^{(i),b}_{\kappa+h|\kappa}(u_j)$, can be obtained by adding a bootstrapped residual, $\widehat e_{\kappa +h|\kappa}^{(i),b}(u_j)$ to the point forecast, $\widehat{\mathcal{X}}^{(i)}_{\kappa+h|\kappa}(u_j)$. The $\widehat e_{\kappa +h|\kappa}^{(i),b}(u_j)$ is generated using sampling with replacement from the in-sample-forecast errors, 
\[
\widehat e_{\delta +h|\delta}^{(i)}(u_j) = \mathcal{X}_{\delta+h}^{(i)}(u_j)-\widehat {\mathcal{X}}_{\delta+h|\delta}^{(i)}(u_j), \qquad \delta = 2,\dots,T-h. 
\]
For a given significance level $\alpha$, the $h$-step-ahead bootstrapped point-wise prediction interval for the $i^\text{th}$ cross section, $[\widehat{\mathcal{X}}^{(i),\text{lb}}_{\kappa+h|\kappa}(u_j), \widehat{\mathcal{X}}^{(i),\text{ub}}_{\kappa+h|\kappa}(u_j)]$, can be calculated point-wise as the $100\times(\alpha/2)^{\text{th}}$ and $100\times(1-\alpha/2)^{\text{th}}$ percentile of the bootstrapped forecasts at each grid point, respectively. The estimation procedure and the forecasting method are summarized in Algorithm 1. 

\begin{algorithm}[!htb]
{\footnotesize{
\DontPrintSemicolon
\SetAlgoLined
 \textbf{Input}: HDFTS $\Big\{\bb{X}_{t}(u) = [\mathcal{X}_{t}^{(1)}, \mathcal{X}_{t}^{(2)}, \ldots, \mathcal{X}_{t}^{(N)}], t = 1, \ldots, \kappa\Big\}$.\\
  \textbf{Estimation Step}:  
 \begin{itemize}
   \item[2.1] For $i\in\{1, \ldots, N\}$, compute the sample mean $\widehat\mu^{(i)} = \frac{1}{\kappa}\sum_{t=1}^{\kappa}\mathcal{X}_t^{(i)}(u)$;
   \item[2.2] For $i\in\{1, \ldots, N\}$, compute the demeaned function $\widehat\mathcal{Y}_t^{(i)} = \mathcal{X}_t^{(i)}(u) - \widehat\mu^{(i)}$;
   \item[2.3] For $i\in\{1, \ldots, N\}$, perform dynamic FPCA to $\widehat\mathbfcal{Y}^{(i)} = [\widehat\mathcal{Y}_1^{(i)}, \widehat\mathcal{Y}_2^{(i)},\ldots, \widehat\mathcal{Y}_{\kappa}^{(i)}]^{\top}$ and obtain the eigenfunctions $\bb{\widehat \lambda^{(i)}(u)} = [\widehat \lambda^{(i)}_1(u), \ldots, \widehat \lambda^{(i)}_{k_i}(u)]$ corresponding to the first $k_i$ largest eigenvalues of the long-run covariance operator of $\widehat\mathbfcal{Y}_t^{(i)}$, where $k_i$ is selected based on~\eqref{eq:6};
   \item[2.4] Choose $k = {\operatorname{max}}\Big\{k_i:1\leq i\leq N\Big\}$, such that the loadings of each cross section are of the same dimension such that $\bb{\Lambda}(u) = [\bb{\widehat \lambda^{(1)}(u)}, \bb{\widehat \lambda^{(2)}(u)}, \ldots, \bb{\widehat \lambda^{(N)}(u)}]^{\top}$;
   \item[2.5] For $t\in\{1, \ldots, \kappa\}$, perform the concurrent functional regression with $\{\widehat\mathcal{Y}_t^{(i)}, i = 1, \ldots, N\}$ as the response and $\{\bb{\widehat \lambda^{(i)}(u)}, i = 1, \ldots, N\}$ as covariates and obtain the coefficient functions $\bb{\widehat{\mathbfcal{F}}}_t(u) = [{\widehat{\mathcal{F}}}_{t,1}(u), \ldots, {\widehat{\mathcal{F}}}_{t,k}(u)]$ and $\bb{\widehat\mathcal{F}}(u) = [\bb{\widehat\mathcal{F}}_1(u)^{\top}, \bb{\widehat\mathcal{F}}_2(u)^{\top}, \ldots, \bb{\widehat\mathcal{F}}_T(u)^{\top}]^{\top}$;
   \item[2.6] For $i, j \in\{1, 
  \ldots, k\}$, compute $\widehat{C}_{h,lj}(u, v) = \frac{1}{T-h}\sum_{t=1}^{T-h}\widehat{\mathcal{F}}_{t, l}(u)\widehat{\mathcal{F}}_{t+h, j}(v)$;
  \item[2.7] For a predetermined $h_0$, compute $\widehat{M}(u,v) = \sum_{h=1}^{h_0}\sum_{l=1}^{k}\sum_{j=1}^{k}\int_{\mathcal{I}}\widehat{C}_{h,lj}(u, z)\widehat{C}_{h,lj}(v, z)dz$;
  \item[2.8] Conduct eigendecomposition on $\widehat{M}(u,v)$ and get $\widehat{\bb{\phi}}(u) = [\widehat{\phi}_1(u), \ldots, \widehat{\phi}_r(u)]$, the eigenfunctions corresponding to the first $r$ largest eigenvalues of $\widehat{M}(u,v)$, where $r$ is selected based on similar idea as in~\eqref{eq:6};
  \item[2.9] For $t\in\{1, \ldots, \kappa\}$, compute $\widehat{\bb{F}}_t = \int\widehat{\bb{\phi}}(u)^{\top}\widehat{\bb{\mathcal{F}}}_t(u)^{\top}du$.
 \end{itemize}
 \textbf{3. Forecasting Step}: 
 \begin{itemize}
   \item[3.1] Obtain the $h$-step ahead forecast of $\{\widehat{\bb{F}}_t , t= 1, \ldots, \kappa\}$, $\widehat{\bb{F}}_{\kappa+h|\kappa}$;
   \item[3.2] Recover the $h$-step-ahead forecast of the $i^\text{th}$ cross section of the HDFTS, $\widehat{\mathcal{X}}^{(i)}_{\kappa+h|\kappa}(u) = \widehat{\mu}^{(i)}(u) + \widehat{\bb{\phi}}(u)\widehat{\bb{F}}_{\kappa+h|\kappa}\widehat{\bb{\lambda}}^{(i)}(u)$. 
 \end{itemize}     
 \KwResult{$\widehat{\mathcal{X}}^{(i)}_{\kappa+h|\kappa}(u)$, the $h$-step-ahead forecast of the $i^\text{th}$ cross section of the HDFTS.}
\caption{\footnotesize{HDFTS forecasting}}}}
\end{algorithm}

\section{Simulation studies}\label{sec:5}

We consider a series of Monte-Carlo simulations to evaluate the finite-sample performance of the proposed method. The data generating process (DGP) is designed to show the proposed method's ability to accurately recover the front factor loading $\bb{\phi}(u)$ and the back factor loading $\bm{\lambda}^{(i)}(u)$ of FTS. 
\begin{enumerate}
\item[1)] The simulated $\{\Y_t^{(i)}(u)\}$ functions are highly positively correlated, corresponding to the strong temporal correlation. This is achieved by adopting VAR models with positive-diagonal coefficient matrices.
\item[2)] The elements in the back factor loading $\bm{\lambda}^{(i)}(u)$ are designed to slightly vary according to $i= 1,\ldots,N$.
\item[3)] The elements of front factor loading $\bb{\phi}(u)$ are selected to be independent of the individual index $i$. This setting is consistent with the practice that one set of homogenous basis functions is used in modelling highly correlated data.
\end{enumerate}

In total, we generate $N$ sets of correlated FTS according to \eqref{eq:1}. In particular, the $t^{\text{th}}$ curve of the $i^{\text{th}}$ FTS is generated from the following model contaminated with measurement error $\epsilon_{t}^{(i)}(u)$
\begin{equation*}
  \mathcal{Y}_{t}^{(i)}(u) =  \bb{\phi}(u)\bb{F}_t [\bb{\lambda}^{(i)}(u)]^{\top} + \epsilon_{t}^{(i)}(u), 
\end{equation*}
where $\bb{F}_t$ is a $2\times 2$ matrix with each row generated using a VAR model of order $1$. Specifically, the first and second rows are generated using VAR(1) model with coefficient matrices 
\[
\begin{pmatrix}
0.7 & 0.2\\0.2 & 0.7
\end{pmatrix},
\qquad
\begin{pmatrix}
0.5 & -0.25\\-1 & 0.5
\end{pmatrix}.
\]
The covariance matrix of the innovations for both VAR(1) models are $\begin{psmallmatrix}1 & 0.5\\0.5 & 1\end{psmallmatrix}$.  The two front-loading curves are $\phi_1(u) = \sin(4\pi u)$ and $\phi_2(u) = \cos(4\pi u)$, respectively, with \{$u = \frac{m}{100} : m = 0,1,\ldots,100$\}. The two back-loading curves for each $i$ are $\lambda_2^{(i)} = \sin(2\pi u + i\pi/4)$ and $\lambda_1^{(i)} = \cos(2\pi u + i\pi/4)$, respectively. The measurement error $\epsilon_{t}^{(i)}(u)$ is generated from independent and identically distributed $\mathcal{N}(0, 0.5^2)$ for all $i$ and $u$. 

The considered DGP involves different combinations of $N$ and $T$. We select a moderate size $T = 20, 40$ and choose $N$ to equal $\frac{T}{2}$, $T$ and $2T$. The pre-determined integer $h_0$ in~\eqref{eq:8} is taken to be one as the $\text{VAR}(1)$ model is considered.

The in-sample model fitting can be evaluated using root mean squared error (RMSE)
\begin{equation*}
  \text{RMSE}(h) = \sqrt{\frac{1}{N\times T\times 101}\sum_{i=1}^{N}\sum_{t=1}^{T}\sum_{m=1}^{101}\big[\mathcal{Y}_t^{(i)}(u_m)-\mathcal{\widehat Y}_t^{(i)}(u_m)\big]^2},
\end{equation*}
where $\mathcal{\widehat Y}_t^{(i)}(u_m)$ represents the estimated function value at grid points $u_m\in [0, 1]$. For each combination of $N$ and $T$, we simulate $100$ replications with pseudo-random seeds and compute the mean of the RMSE values. 

Table~\ref{tab:1} displays the model fitting performance. For each $N$, when $T$ increases, the estimation error is smaller. For each $T$, as $N$ increases, the estimation error is stable, and the proportion of the corrected estimated number of factors significantly increases, which indicates that our model can handle high-dimensional cases when $N>T$.
\begin{table}[!htb] 
\centering 
\caption{Estimation performance for different combinations of $N$ and $T$. RMSE measures the model fitting, and $\widehat r$ and $\widehat k$ measure the proportion of the correctly estimated number of front- and back-loading curves.}\label{tab:1} 
    \setlength{\tabcolsep}{46pt}
   \renewcommand{\arraystretch}{1.5}
\begin{tabular}{@{}llrrr@{}} 
\toprule
$T$ & $N$ & RMSE & $\widehat r = 2$ \text{or} $3$  & $\widehat k = 2$ \\
\midrule 
\multirow{3}{*}{40} & 20 & $1.12$ & $0.66$  & $0.93$ \\
&  $40$& $0.99$ & $0.73$  & $0.96$ \\  
& $80$ & $1.07$ & $0.82$  & $0.98$ \\ 
\bottomrule
\end{tabular} 
\end{table} 

We also compare the forecast performance of the proposed model with the univariate FTS forecasting method of \cite{HyndmanUllah2007}. The simulated data is split into two sets in evaluating the forecasting performance, that is, the training set and the testing set. Different models are fitted with the training dataset, and forecasts are made based on different models. Then, the forecasts are compared with the actual observations in the testing set. In this case, we set the starting size of the training set and testing set as $\frac{3}{4}\times T$ and $\frac{1}{4}\times T$, respectively. We use expanding window analysis, where we increase the training size in each iteration by one and refit the models. We obtain the one- to ten -step-ahead forecasts and compute our forecasts' root mean square forecast error (RMSFE). 
\begin{equation*}
  \text{RMSFE}(h) = \sqrt{\frac{1}{N\times (\frac{1}{4}\times T-h+1)\times 101}\sum_{i=1}^{N}\sum_{\kappa=\frac{3}{4}\times T}^{T-h}\sum_{m=1}^{101}\big[\mathcal{Y}_{\kappa+h}^{(i)}(u_m)-\mathcal{\widehat Y}_{\kappa+h|\kappa}^{(i)}(u_m)\big]^2},
\end{equation*}
where $\kappa$ is the size of the data used to generate forecast and $\mathcal{\widehat Y}_{\kappa+h|\kappa}^{(i)}(u_m)$ represents the forecast values at grid points $u_m$.  

For each combination of $N$ and $T$, we simulate $100$ replications and compute the mean of the RMSFE values of different models. Table~\ref{tab:2} shows the mean RMSFE values for different combinations of $N$ and $T$ for both models. The bold entries highlight the method that produces the most accurate forecast. The proposed model produces more accurate forecasts for most of the cases for all the combinations of $N$ and $T$, demonstrating the proposed model's superiority in forecasting HDFTS. 

\begin{table}[!htb] \centering 
\caption{The mean of RMSFE values of different models for different combinations of $N$ and $T$, ``FDFM'' represents the proposed model and "UFTS" represents the univariate FTS forecasting model.} 
  \label{tab:2} 
    \setlength{\tabcolsep}{21pt}
   \renewcommand{\arraystretch}{1.5}
\begin{tabular}{@{}llrrrrrr@{}} 
\toprule
& & \multicolumn{2}{c}{$N=20$} & \multicolumn{2}{c}{$N=40$} &\multicolumn{2}{c}{$N=80$}\\
\cmidrule{3-8}
$T$ & $h$ & UFTS & FDFM & UFTS & FDFM & UFTS & FDFM\\
\cmidrule{1-8}
\multirow{5}{*}{$40$}& $1$ & $2.189$ & $\textbf{1.381}$ & $2.647$ & $\textbf{1.478}$ & $2.758$ & $\textbf{1.555}$\\
& $2$ & $2.276$ & $\textbf{1.661}$ & $2.743$ & $\textbf{1.862}$ & $2.856$ & $\textbf{1.958}$ \\
& $3$ & $2.362$ & $\textbf{1.858}$ & $2.841$ & $\textbf{2.141}$ & $2.977$ & $\textbf{2.223}$ \\
& $4$ & $2.447$ & $\textbf{2.091}$ & $2.956$ & $\textbf{2.301}$ & $3.008$ & $\textbf{2.356}$ \\
& $5$ & $2.493$ & $\textbf{2.202}$ & $3.131$ & $\textbf{2.530}$ & $3.184$ & $\textbf{2.589}$ \\
\bottomrule
\end{tabular} 
\end{table} 

\section{Empirical studies}\label{sec:6}

We apply the proposed method to forecast sub-national mortality rates in Japan \citep{jmd}. In particular, we consider the age-specific mortality rates of $47$ prefectures of Japan from $1975$ to $2021$. We pick this time interval to ensure the number of observation years is less than the number of prefectures, ensuring the ``curse of dimensionality" challenges are preserved in the empirical data. 

We evaluate and compare the forecasting accuracy of the proposed method with several competing methods. Further, we estimate life annuity pricing to demonstrate the impact of the forecasting improvement from the proposed method. 

\subsection{Japanese subnational age-specific mortality}

\begin{figure}[!htb]
\centering
\subfloat[Female smoothed mortality rates in Tokyo]{{\includegraphics[width=8.4cm]{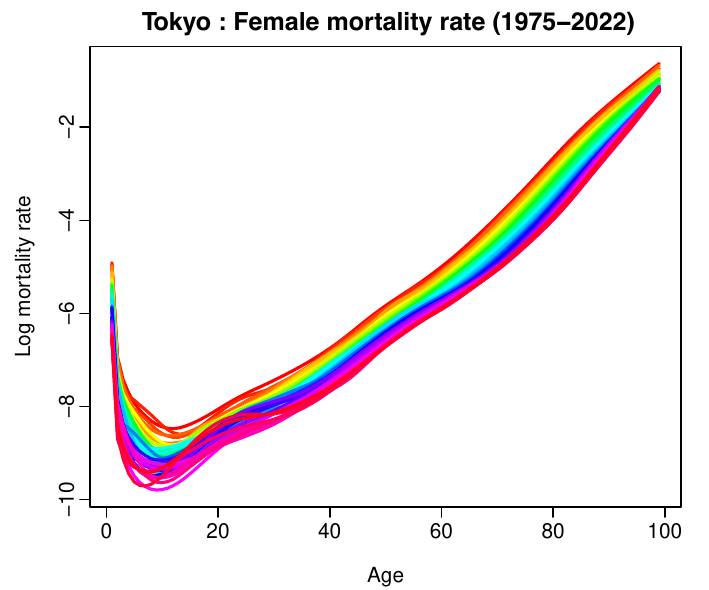} }}
\qquad
\subfloat[Female smoothed mortality rates in Nagano]{{\includegraphics[width=8.4cm]{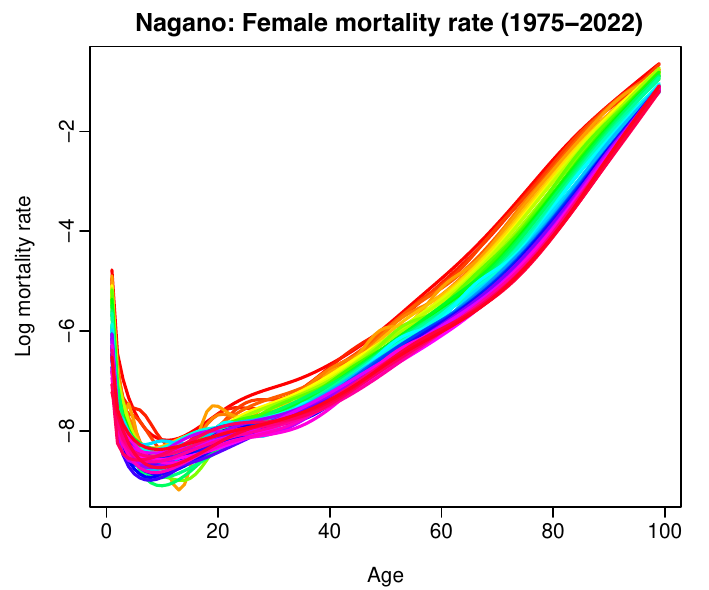} }}
    \qquad
    \subfloat[Female smoothed mortality rates in Kyoto]{{\includegraphics[width=8.4cm]{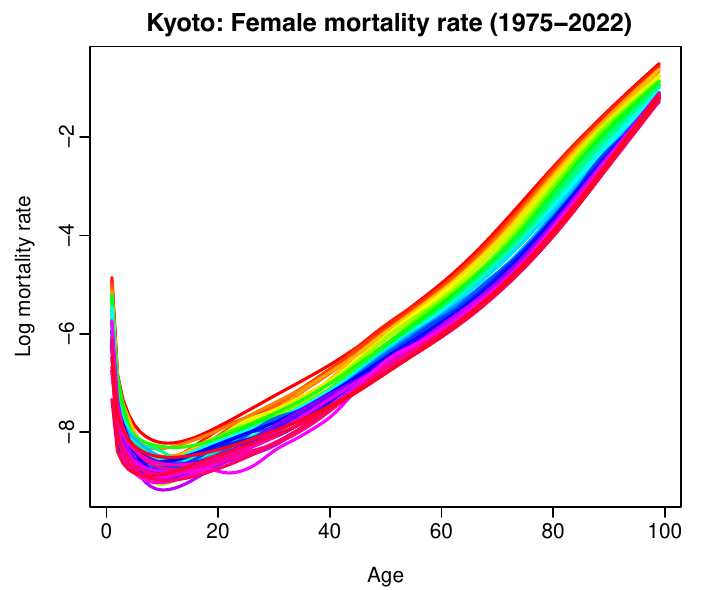} }}
    \qquad
    \subfloat[Female smoothed mortality rates in Osaka]{{\includegraphics[width=8.4cm]{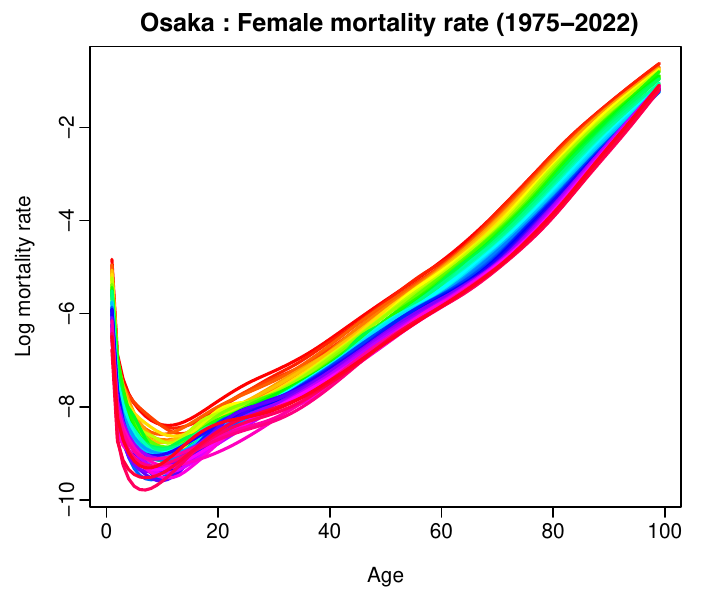} }}
    \caption{Functional time series displays for smoothed age-specific mortality rates of selected prefectures in Japan.}%
    \label{fig:1}
\end{figure}

Figure~\ref{fig:1} depicts the rainbow plots of the logarithm of the smoothed female mortality rates for four prefectures of Japan, namely, Tokyo, Nagano, Kyoto and Osaka, from $1975$ to $2022$. The rainbow plots of \cite{hyndman2010rainbow} can visualize a time ordering of functions. The colour order of the rainbow reflects the time ordering of the functional observations. Functions from the distant past are depicted in red, whereas more recent functions are shown in purple. 

The weighted penalized regression splines method is applied to convert the observed data into continuous functions. Due to the sparsity of data for very high age group, we have aggregated both female and male mortality for ages over $98$, hence the number of grid points on the mortality curves is $99$.

Despite the general shapes of mortality rates for these four prefectures being similar, we may observe significant differences in the troughs and peaks and the overlaps of the curves. This indicates the mortality rates of different prefectures contain heterogeneity. To achieve improved forecasts of the subnational mortality rates, we need to extract the common features and use these common features to produce forecasts. This could be achieved by separating the homogeneity and heterogeneity of mortality rates of all prefectures. From~\eqref{eq:4}, the heterogeneity of mortality rates of each prefecture is reflected in $\bb{\Lambda}(u)$, the $N\times k$ functional factor loading matrix, since each prefecture has a different set of functional factor loading. The term $\bb{\phi}(u)\bb{F}_t$ represents the homogeneity, which is common to all FTS.

\begin{figure}[!htb]
\centering
\subfloat[First functional factor loading]
{{\includegraphics[width=8.8cm]{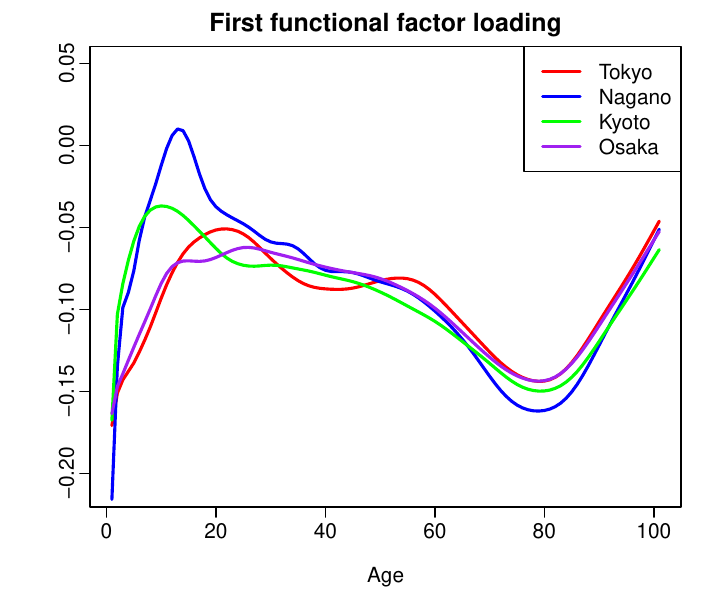}}}
\quad
\subfloat[Second functional factor loadingo]
{{\includegraphics[width=8.8cm]{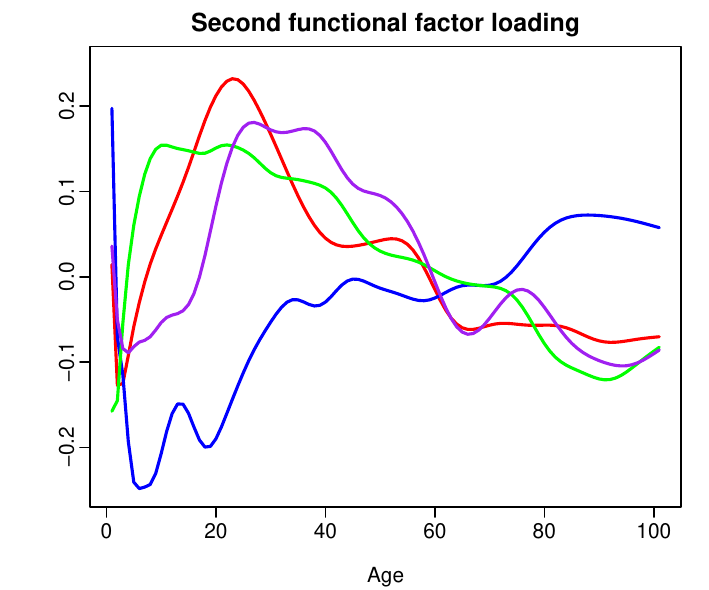}}}
\\
\subfloat[Third functional factor loading]
{{\includegraphics[width=8.8cm]{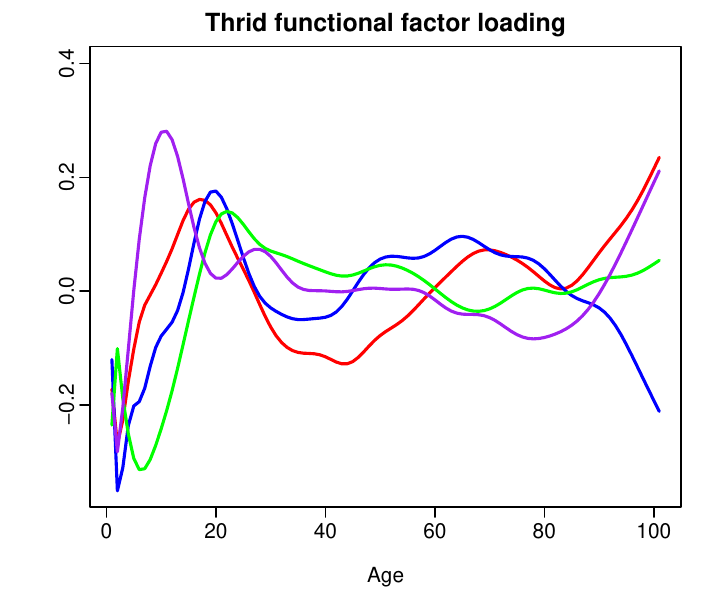}}}
\quad
\subfloat[Fourth functional factor loading]
{{\includegraphics[width=8.8cm]{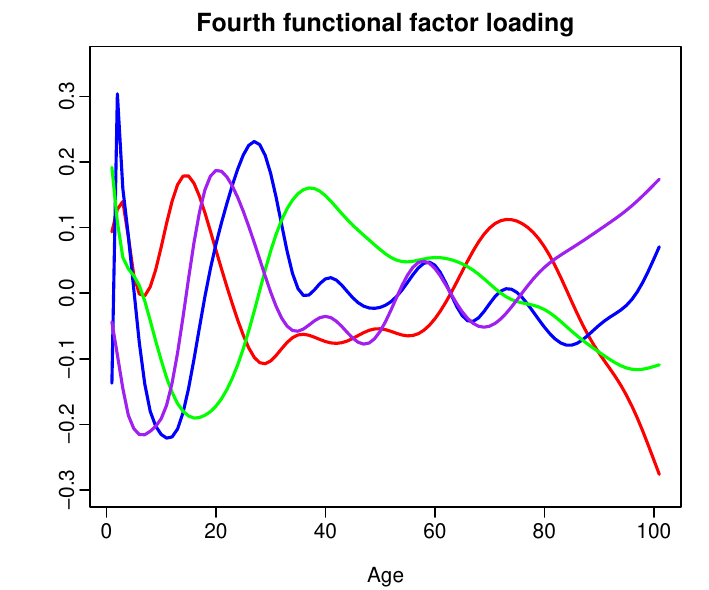}}}
\caption{The first four functional factor loadings for four prefectures of Japan, namely Tokyo, Nagano, Kyoto and Osaka.}%
\label{fig:2}
\end{figure}

Figure~\ref{fig:2} shows the first four functional factor loadings of mortality rates of four chosen prefectures. All these four-factor loadings are different for these four prefectures. This represents the heterogeneity in the subnational mortality rates with different troughs, peaks and overlaps of the mortality rate curves.

\begin{figure}[!htb]
\centering
\subfloat[First set of common FTS]
{{\includegraphics[width=8.8cm]{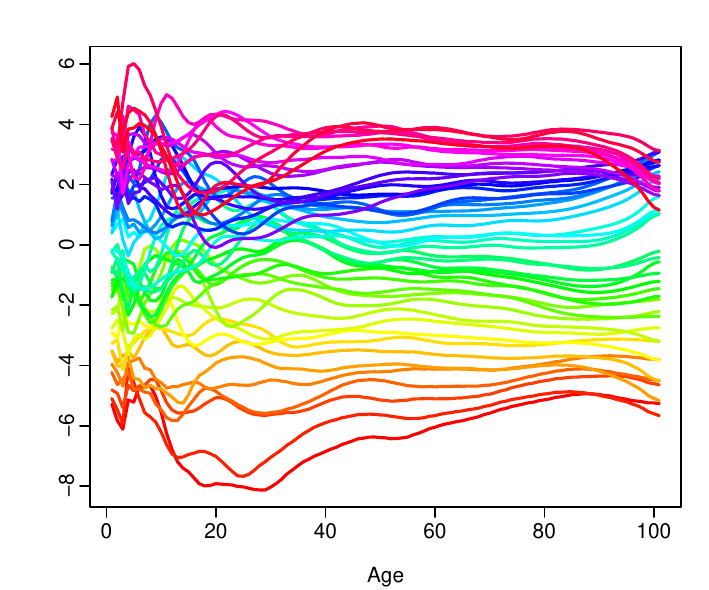}}}
\quad
\subfloat[Second set of common FTS]
{{\includegraphics[width=8.8cm]{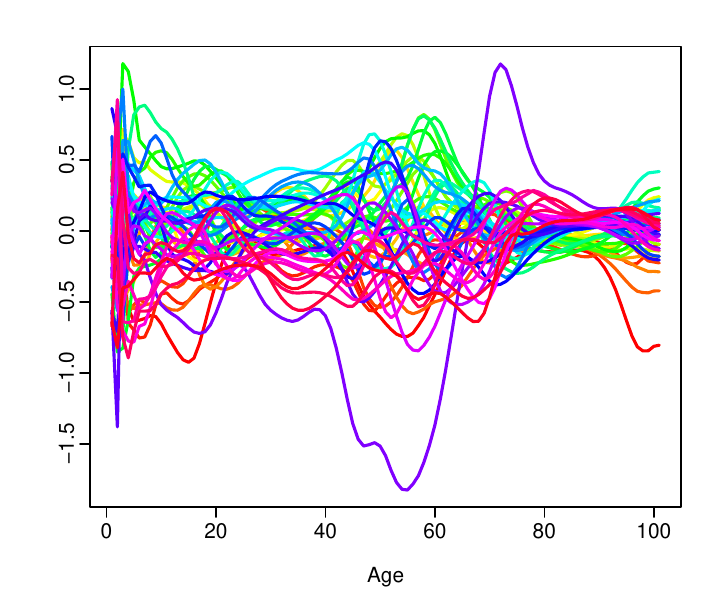}}}
\\
\subfloat[Third set of common FTS]
{{\includegraphics[width=8.8cm]{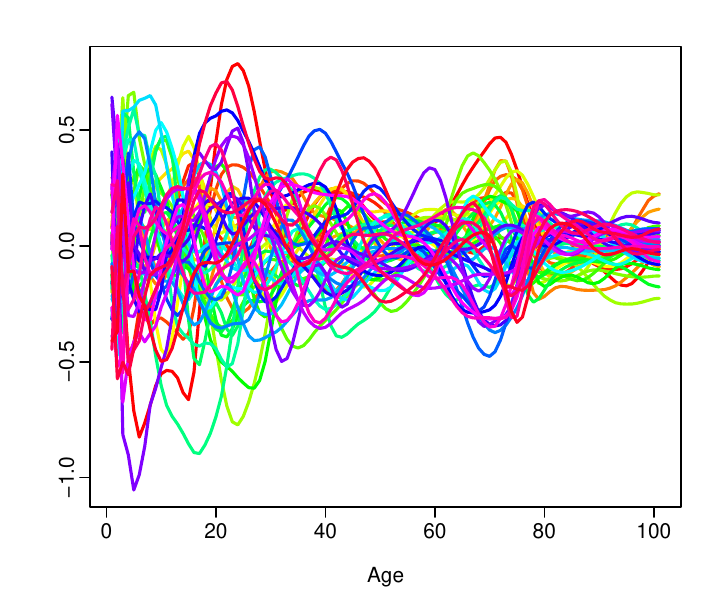}}}
\quad
\subfloat[Fourth set of common FTS]
{{\includegraphics[width=8.8cm]{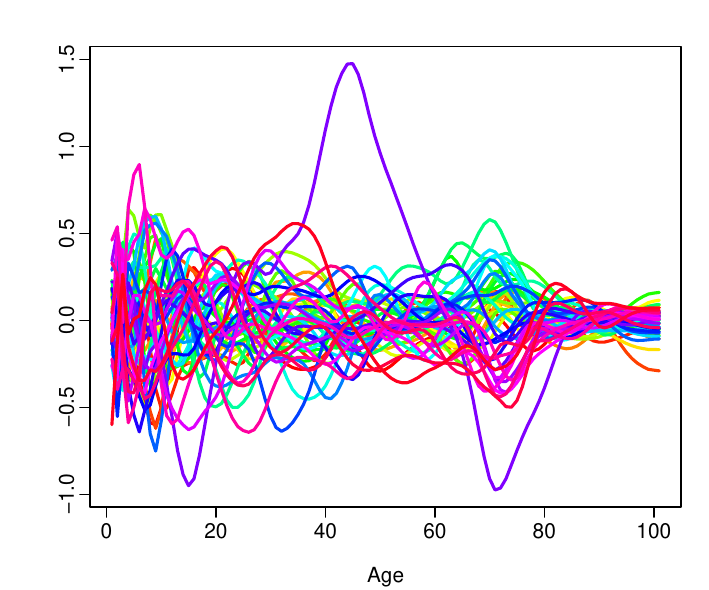}}}
\caption{First four sets of common FTS of selected prefectures (Tokyo, Nagano, Kyoto and Osaka) in Japan.}\label{fig:3}
\end{figure}

After separating the heterogeneity, we extract the common features in Figure~\ref{fig:3}, which are common to all mortality rate curves. This reflects the general common shapes and colour ordering of the mortality rates of all prefectures of Japan. By extracting the common features, we have reduced the dimension of the FTS of $47$ prefectures of Japan into smaller sets of FTS. The temporal information within the original HDFTS is preserved in the fixed-dimensional FTS.

\begin{figure}[!htb]
\centering
\subfloat[Common factor loadings for female]
{{\includegraphics[width=8.8cm]{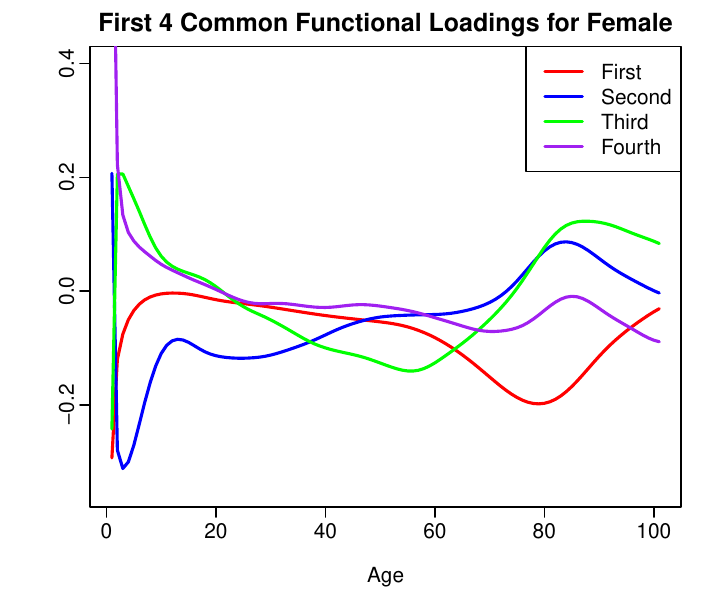}}}
\quad
\subfloat[Common factor loadings for male]
{{\includegraphics[width=8.8cm]{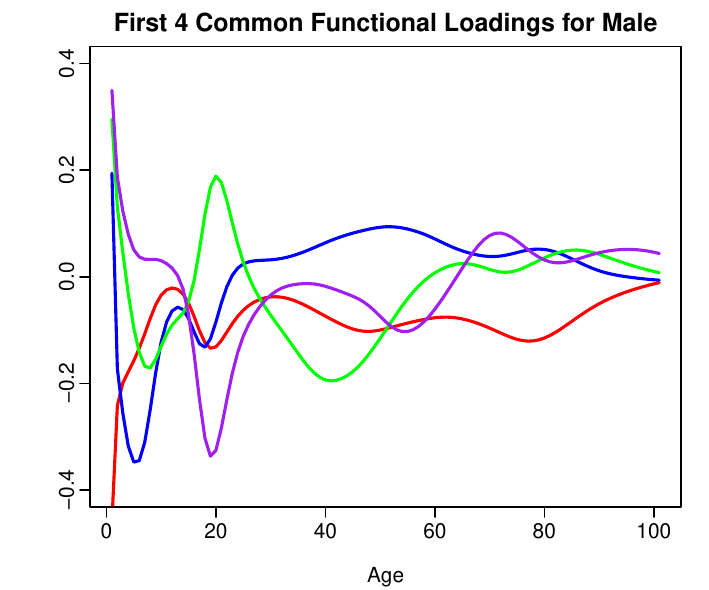}}}
\caption{The first four common factor loadings for female and male data}\label{fig:4}
\end{figure}

In addition, the first four common functional factor loadings of both genders are depicted in Figure~\ref{fig:4}. Despite the general shape being similar for males and females, the common functional loadings for males contains more troughs and peaks, indicating that male mortality rates are more volatile than those of females. 

Table~\ref{tab:5} presents the first six eigenvalues for the auto-cross-covariance function of the common FTS and the long-run-covariance function of the FTS of four selected prefectures for both genders. The eigenvalues are normalized so that the sum of all eigenvalues adds up to one to reflect the percentage of total variation explained. We observe that the first few factors would explain most of the variations in the data.
\begin{table}[!htb] 
\centering 
  \caption{The first six normalized eigenvalues for the auto-cross-covariance function of the common FTS and the long-run-covariance function of four selected prefectures for both genders are reported. The eigenvalues for the auto-cross-covariance function of the common FTS are labelled as ``Common'', and the eigenvalues for the long-run-covariance function of the FTS are labelled as ``Specific".} 
  \label{tab:5} 
\setlength{\tabcolsep}{3.5pt}
\renewcommand{\arraystretch}{1.5}
\begin{tabular}{@{\extracolsep{0pt}} lrrrrrrrrrr@{}} 
\toprule
& \multicolumn{5}{c}{Female}& \multicolumn{5}{c}{Male}\\
\cmidrule{2-6}\cmidrule{7-11}
 & \multirow{2}{*}{Common} & \multicolumn{4}{c}{Specific}& \multirow{2}{*}{Common}& \multicolumn{4}{c}{Specific}\\
  \cmidrule{3-6}  \cmidrule{8-11}
Eigenvalue &  & Tokyo & Nagano & Kyoto & Osaka &  & Tokyo & Nagano & Kyoto & Osaka  \\ 
 \midrule
1\textsuperscript{st} & $0.901$ & $0.973$ & $0.957$ & $0.961$ & $0.971$ & $0.873$ & $0.959$ & $0.933$ & $0.937$ & $0.956$ \\ 
2\textsuperscript{nd} & $0.055$ & $0.018$ & $0.026$ & $0.018$ & $0.016$ & $0.048$ & $0.019$ & $0.024$ & $0.029$ & $0.024$ \\ 
3\textsuperscript{rd} & $0.023$ & $0.004$ & $0.007$ & $0.010$ & $0.006$ & $0.032$ & $0.012$ & $0.015$ & $0.013$ & $0.008$ \\ 
4\textsuperscript{th} & $0.008$ & $0.002$ & $0.004$ & $0.005$ & $0.004$ & $0.021$ & $0.004$ & $0.009$ & $0.008$ & $0.005$ \\ 
5\textsuperscript{th} & $0.007$ & $0.001$ & $0.002$ & $0.003$ & $0.002$ & $0.009$ & $0.002$ & $0.005$ & $0.005$ & $0.003$ \\ 
6\textsuperscript{th} & $0.003$ & $0.001$ & $0.001$ & $0.002$ & $0.001$ & $0.006$ & $0.001$ & $0.005$ & $0.003$ & $0.002$ \\ 
\midrule
Sum & $0.997$ & $0.999$ & $0.997$ & $0.999$ & $1.000$ & $0.989$ & $0.997$ & $0.991$ & $0.995$ & $0.998$ \\ 
\bottomrule
\end{tabular} 
\end{table} 

The numbers of factors that we select for the common and specific FTS are four and six, respectively, for females and seven and eight, respectively, for males. With these selections of the number of factors, we ensure that the factors explain $99$\% of the variation in the data. Moreover, the percentages of total variation explained by the first factor for males are smaller than those for females, and this also explains that the data for males are more volatile. 

\subsection{Forecast evaluation}

We consider the Japanese sub-national mortality data between 1975--2022 in the empirical study. To assess model and parameter stabilities over time, we adopt an expanding window analysis \citep[see][Chapter 9 for details]{zivot2006modeling} and produce forecasts in iterations. Specifically, we initiate the process by fitting the proposed HDFTS method to the first 38 years of observations between 1975--2012 and generating one- to 10-step-ahead point forecasts. Through an expanding window approach, we re-estimate the parameters of the considered model using the first 39 mortality curves from 1975 to 2013 and make one- to nine-step-ahead forecasts. The process is iterated with the sample size increased by one year until the end of the observation period. This process produces ten one-step-ahead forecasts, nine two-step-ahead forecasts, and so on, up to one 10-step-ahead forecast. 



We compare the proposed method with the HDFTS studied in \cite{gao2018high}, the high-dimensional functional factor model \citep[HDFFM,][]{nisol2019high}, the two-step high-dimensional functional time series model \citep[TSHDFTS][]{chang2024} the factor models for matrix-valued high-dimensional time series \citep[MFM,][]{wang2019factor} and the tensor decomposition forecast methods (Canonical Polyadic Decomposition (CPD) and Tucker of \cite{dong2020multi}).

\subsubsection{Point forecast accuracy}

We use the root mean square forecast error (RMSFE) to evaluate the point forecast accuracy. The RMSFE measures how close the forecast results are to the actual values of the data under forecast. The RMSFE for the $h$-step-ahead forecasts of the $i^\text{th}$ prefecture for $i \in \{1, \ldots, 47\}$ can be written as 
\begin{equation*}
  \text{RMSFE}^{(i)}(h) = \sqrt{\frac{1}{J\times(T-n-h+1)}\sum_{\kappa=n}^{T-h}\sum_{m=1}^{101}\big[\mathcal{X}_{\kappa+h}^{(i)}(u_m)-\widehat{\mathcal{X}}_{\kappa+h|\kappa}^{(i)}(u_m)\big]^2},
\end{equation*}
where $\kappa$ is the number of observations used in generating the point forecasts, $\mathcal{X}_{\kappa+h}^{(i)}(u_j)$ is the actual value of the ${(\kappa+h)}^{\textsuperscript{th}}$ curve of the $i^\text{th}$ prefecture, $\widehat{\mathcal{X}}_{\kappa+h}^{(i)}(u_j)$ is the $h$-step-ahead point forecast based on the first $\kappa$ observations, and $J$ is the number of grid points on each curve. We use the average RMSFE for the $h$-step-ahead forecasts across all prefectures to reflect the accuracy of the $h$-step-ahead point forecasts:
\begin{equation*}
  \overline{\text{RMSFE}}(h) = \frac{1}{N}\sum_{i=1}^{N}\text{RMSFE}^{(i)}(h).
\end{equation*}
Table~\ref{tab:3} tabulates the average RMSFE values across all prefectures for one-step ahead to $10$-step ahead forecasts of different forecast methods. The bold entries highlight the method that produces the most accurate point forecast. Forecasts based on the proposed method have the smallest average RMSFE for both male and female mortality, which indicates that the proposed method outperforms other methods in terms of point forecast. This implies that the proposed model extracts the most information contained in the HDFTS. 

\begin{small}
\begin{centering}
\renewcommand{\arraystretch}{0.7}
\setlength{\tabcolsep}{10pt}
\begin{longtable}{@{\extracolsep{5pt}}llccccccc@{}}
\caption{Average RMSE values ($\times 100$) in the holdout sample based on various forecasting methods are presented. Forecasts based on the proposed method are labelled as ``FDFM'', forecasts based on high-dimensional functional time series are labelled as ``HDFTS'', forecasts based on the two-step high-dimensional functional time series model are labelled as ``TSHDFTS", forecasts based on high-dimensional functional factor models are labelled as ``HDFFM'', forecasts based on matrix factor models are labelled as ``MFM'', forecasts based on CPD tensor decompositions are labelled as ``CPD'' and forecasts based on Tucker tensor decompositions are labelled as ``Tucker''.}\\
\toprule
\label{tab:3}
Sex & $h$ & FDFM & TSHDFTS & HDFTS & HDFFM & MFM & CPD & Tucker \\
\midrule
Female & $1$ & $\textBF{0.641}$ & $0.723$ & $2.165$ & $2.272$ & $1.464$ & $0.755$ & $0.741$  \\ 
& $2$ & $\textBF{0.653}$ & $0.737$ & $2.225$ & $2.208$ & $1.496$ & $0.745$ & $0.737$ \\ 
& $3$ & $\textBF{0.655}$ & $0.755$ & $2.268$ & $2.261$ & $1.518$ & $0.730$ & $0.703$  \\ 
& $4$ & $\textBF{0.715}$ & $0.772$ & $2.319$ & $2.362$ & $1.537$ & $0.761$ & $0.729$  \\ 
& $5$ & $\textBF{0.691}$ & $0.801$ & $2.360$ & $2.397$ & $1.553$ & $0.770$ & $0.774$  \\ 
& $6$ & $\textBF{0.723}$ & $0.787$ & $2.420$ & $2.499$ & $1.582$ & $0.739$ & $0.739$  \\ 
& $7$ & $0.815$ & $0.797$ & $2.470$ & $2.513$ & $1.601$ & $0.741$ & $\textBF{0.740}$  \\ 
& $8$ & $0.817$ & $0.842$ & $2.516$ & $2.698$ & $1.617$ & $0.810$ & $\textBF{0.809}$  \\ 
& $9$ & $\textBF{0.666}$ & $0.924$ & $2.413$ & $2.921$ & $1.488$ & $0.840$ & $0.833$  \\ 
& $10$ & $\textBF{0.722}$ & $1.141$ & $2.199$ & $3.296$ & $1.261$ & $0.839$ & $0.857$  \\ 
\cmidrule{2-9}
& Mean & $\textBF{0.710}$ & $0.828$ & $2.336$ & $2.543$ & $1.512$ & $0.773$ & $0.776$\\
\midrule
Male & $1$ & $\textBF{0.998}$ & $2.444$ & $2.119$ & $2.627$ & $2.970$ & $2.703$ & $2.730$ \\ 
& $2$ & $\textBF{1.004}$ & $2.516$ & $2.172$ & $2.725$ & $2.866$ & $2.7663$ & $2.790$ \\ 
& $3$ & $\textBF{1.030}$ & $2.536$ & $2.218$ & $2.772$ & $2.836$ & $2.783$ & $2.821$ \\ 
& $4$ & $\textBF{1.053}$ & $2.642$ & $2.258$ & $2.637$ & $2.911$ & $2.795$ & $2.810$ \\ 
& $5$  & $\textBF{1.086}$ & $2.850$ & $2.304$ & $2.508$ & $2.794$ & $2.812$ & $2.842$ \\ 
& $6$  & $\textBF{1.093}$ & $3.006$ & $2.372$ & $2.422$ & $2.917$ & $2.841$ & $2.869$ \\ 
& $7$  & $\textBF{1.112}$ & $3.161$ & $2.432$ & $2.484$ & $3.149$ & $2.865$ & $2.905$ \\ 
& $8$  & $\textBF{1.138}$ & $2.634$ & $2.473$ & $2.496$ & $2.937$ & $2.913$ & $2.954$ \\ 
& $9$  & $\textBF{1.290}$ & $1.601$ & $2.322$ & $2.358$ & $2.798$ & $2.892$ & $2.952$ \\ 
& $10$  & $\textBF{1.578}$ & $2.119$ & $2.025$ & $2.052$ & $2.968$ & $2.682$ & $2.756$ \\ 
\cmidrule{2-9}
& Mean  & $\textBF{1.138}$ & $2.551$ & $2.270$ & $2.508$ & $2.915$ & $2.805$ & $2.843$ \\ 
\bottomrule
\end{longtable}
\end{centering}
\end{small}
\subsubsection{Interval forecast accuracy}

We use the interval scoring rule of \cite{gneiting2007strictly} and \cite{GK14} to evaluate point-wise interval forecast accuracy. The interval score for the point-wise prediction interval of the $i^\text{th}$ prefecture at time point $u_{j}$ is 
\begin{equation*}
\begin{split}
S_\alpha\Big[\widehat\mathcal{X}_{\kappa+h|\kappa}^{(i),\text{lb}}(u_m), \widehat\mathcal{X}_{\kappa+h|\kappa}^{(i),\text{ub}}(u_m); \mathcal{X}_{\kappa+h}^{(i)}(u_m)\Big] & = \Big[\widehat\mathcal{X}_{\kappa+h|\kappa}^{(i),\text{ub}}(u_m) - \widehat\mathcal{X}_{\kappa+h|\kappa}^{(i),\text{lb}}(u_m)\Big] \\
  & + \frac{2}{\alpha}\Big[\widehat\mathcal{X}_{\kappa+h|\kappa}^{(i),\text{lb}}(u_m) - \mathcal{X}_{\kappa+h}^{(i)}(u_m)\Big]\mathbbm{1}\Big\{ \mathcal{X}_{\kappa+h}^{(i)}(u_m)<\widehat\mathcal{X}_{\kappa+h|\kappa}^{(i),\text{lb}}(u_m)\Big\}\\
  & + \frac{2}{\alpha}\Big[\mathcal{X}_{\kappa+h}^{(i)}(u_m) -\widehat\mathcal{X}_{\kappa+h|\kappa}^{(i),\text{ub}}(u_m)\Big]\mathbbm{1}\Big\{ \mathcal{X}_{\kappa+h}^{(i)}(u_m)>\widehat\mathcal{X}_{\kappa+h|\kappa}^{(i),\text{ub}}(u_m)\Big\},
\end{split}
\end{equation*}
where the level of significance $\alpha$ can be chosen conventionally as $0.2$. It is not difficult to find that the smaller the interval score, the more accurate the interval forecast. An optimal (which is also minimal) interval score value can be achieved if $\mathcal{X}_{\kappa+h}^{(i)}(u_j)$ lies between $\widehat\mathcal{X}_{\kappa+h|\kappa}^{(i)\text{lb}}(u_j)$ and $\widehat\mathcal{X}_{\kappa+h|\kappa}^{(i)\text{ub}}(u_j)$. Then the mean interval score for the $h$-step-ahead forecasts of the $i^\text{th}$ prefecture can be written as
\begin{equation*}
  \overline S_\alpha^{(i)}(h) = \frac{1}{101\times(T-n-h+1)}\sum_{\kappa=n}^{T-h}\sum_{m=1}^{101}S_\alpha\Big[\widehat\mathcal{X}_{\kappa+h|\kappa}^{(i),\text{lb}}(u_m), \widehat\mathcal{X}_{\kappa+h|\kappa}^{(i),\text{ub}}(u_m); \mathcal{X}_{\kappa+h}^{(i)}(u_m)\Big].
\end{equation*}
We use the average mean interval score for the $h$-step-ahead forecasts across all prefectures to reflect the accuracy of the $h$-step-ahead point-wise prediction interval:
\begin{equation*}
  \overline{S}_{\alpha}(h) = \frac{1}{N}\sum_{i=1}^{N}\overline S_\alpha^{(i)}(h).
\end{equation*}

\begin{figure}[!htb]
\centering
\subfloat[Mean interval scores of various forecasting methods for female mortality]{{\includegraphics[width=8cm]{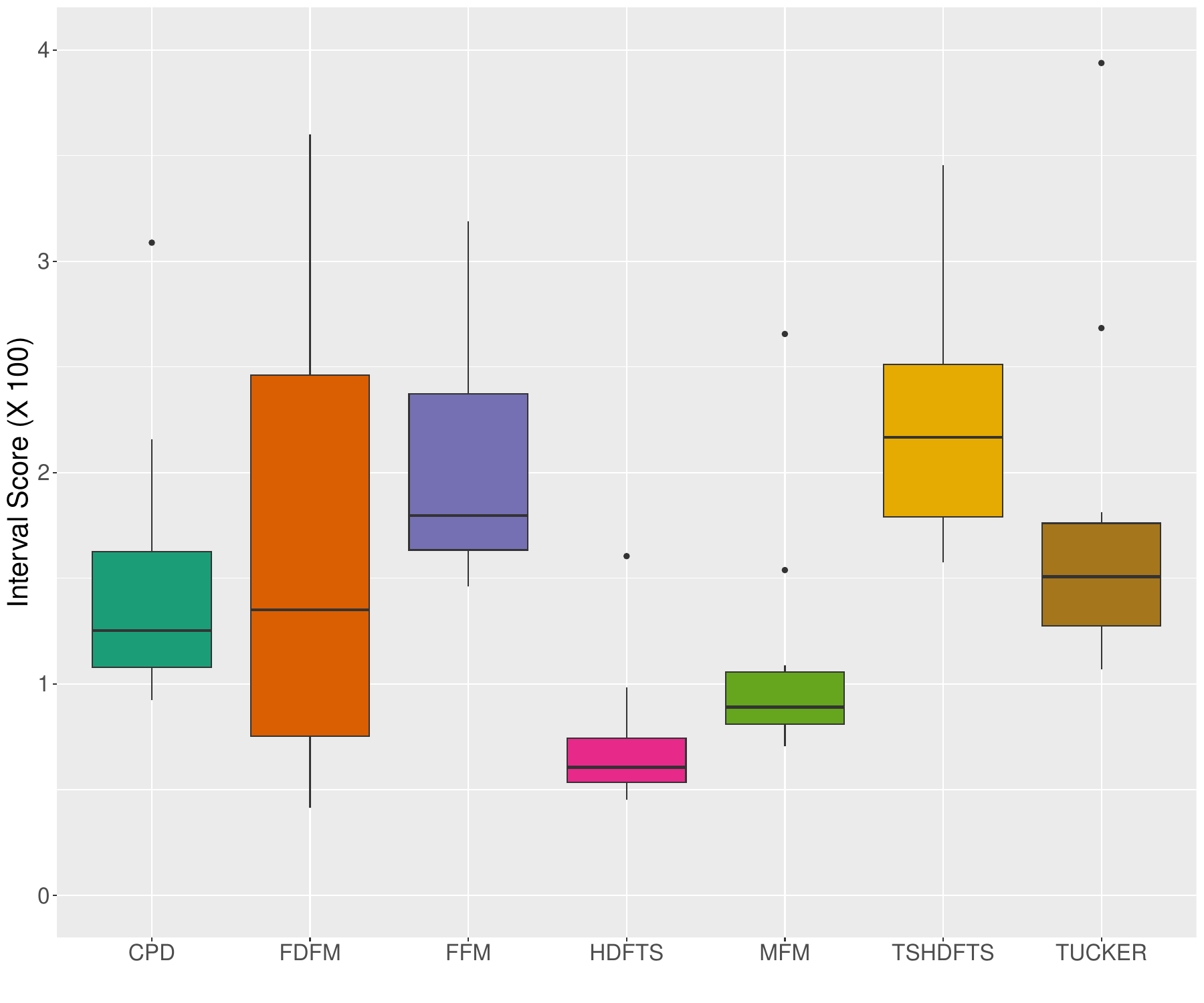}}}
\qquad
\subfloat[Mean interval scores of various forecasting methods for male mortality]{{\includegraphics[width=8cm]{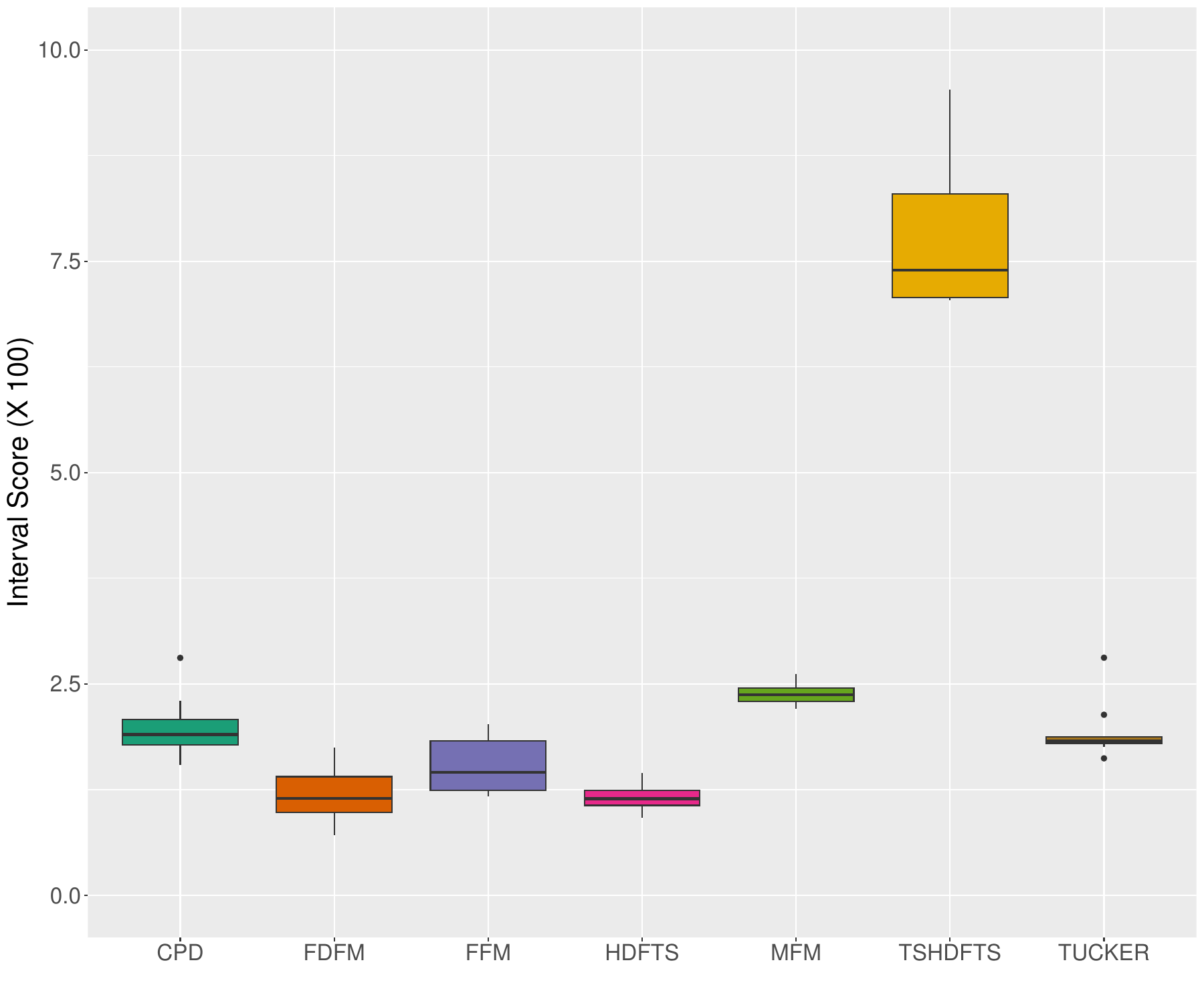} }}
\caption{Average interval score values ($\times 100$) in the holdout sample of various forecasting methods}\label{fig:4_2}
\end{figure}

Figure~\ref{fig:4_2} shows the side-by-side box plots of average mean interval score values across all prefectures for one-step ahead to $10$-step ahead forecasts of different forecast methods. From the box plots, despite HDFTS performing better for female mortality, the proposed method performs relatively the same as HDFTS, which indicates the proposed method could generate relatively smaller average mean interval score values for some of the forecast horizons. This demonstrates that the proposed method is capable of producing more accurate interval forecasts. 

After comparing the point and interval forecast results of various methods, we found that our proposed method outperforms other competitive methods in forecasting the HDFTS. This indicates that the proposed model's dimension reduction process efficiently captures both the serial dependence and the correlation among cross sections in the HDFTS. 

\subsubsection{Computation Speed}\label{sec:7.2.3}

We also compare the computational speed of various high-dimensional mortality forecasting methods. The methods shown in Table~\ref{tabnew} are applied to the same training dataset sequentially to obtain one-step-ahead point forecasts, with their computation time in seconds recorded. It is worth noting that CPD's fast computation speed is due to the time efficiency of Tucker's decomposition embedded in the algorithm. Our proposed method takes 50\% -- 90\% more time to implement compared to other functional data analysis approaches shown in the table. However, the superior forecasting accuracy of the proposed FDFM method justifies the additional computation time. The TSHDFTS method of \cite{chang2024} selects parameters by cross-validations, which requires significant computation time in implementation. For this reason, we left the method out of the comparison presented in the table.

\begin{table}[!htb] 
\centering 
\caption{Computational time (in seconds) of the considered high-dimensional mortality forecasting methods.} \label{tabnew} 
\setlength{\tabcolsep}{40pt}
\renewcommand{\arraystretch}{1.5}
\begin{tabular}{@{}llr@{}} 
\toprule
 & Model & Computation Time (seconds) \\
\midrule
Our method & FDFM & 187.12  \\
\cite{gao2018high} & HDFTS &  68.95 \\
\cite{wang2019factor} & MFM &  95.94 \\
\cite{dong2020multi} & CPD &  0.20 \\
\cite{dong2020multi} & Tucker &  0.39 \\
\cite{nisol2019high} & HDFFM &  108.14\\
\bottomrule
\end{tabular} 
\end{table} 

\subsection{Life annuity Pricing} \label{sec:6.3}

Life annuities are one of the typical longevity insurance products for retirees, helping them finance their retirements. Rapid improvements in mortality have exposed life insurers to longevity risk \citep{ngai2011longevity}. Accurate mortality forecasts could enable life insurers to manage longevity risk effectively without holding excessive levels of capital. To illustrate the impact of these forecasting improvements, we use the mortality forecasts to price life annuities, that is, the amount of money an individual pays to a life insurer in return for annual payments after retirement until death. We compare the present values of the life annuities, that is, the amount of capital the life insurer should reserve, based on mortality forecasts from different methods. 

In comparing the value of a life annuity, we calculate the present value (PV) of the life annuity with \$1 annual payments. More specifically, the price of a life annuity for an individual aged $x$ at year $t$ is the PV of the annual payments of \$1 that the individual receives after retirement until death or a pre-agreed age (whichever occurs first). The retirement age is set to be $65$, and the pre-agreed age at which the annuity terminates is assumed to be $90$ \citep{he2021data}. Then the price of the annuity can be calculated as follows: 
\begin{align*}
  \text{PV}_{x, t} = \left\{ \begin{array}{ll}
                \sum_{n=1}^{90-x} \frac{_np_{x,t}}{(1+i)^n}, & x \geq 65 \\ 
                \\
                \sum_{n=1}^{25} \frac{_np_{65,t+(65-x)}}{(1+i)^{n+(65-x)}}, &  x < 65,\\
                \end{array} \right.
\end{align*}
where $\text{PV}_{x, t}$ is the PV of the life annuity for an individual aged $x$ at year $t$, $_np_{x,t}$ is the survival probability for an individual aged $x$ at year $t$ to survive after $n$ years, and $i$ is the interest rate used for discounting. An individual older than $ 65$-year-old receives payment for each year of survival, and for an individual younger than $ 65$-year-old, the annuity is deferred with the first payment paid out at the year that they survive their $66^{\text{th}}$ birthday. 

To demonstrate the financial impact of the forecasting improvement in mortality, we compare annuity prices based on mortality forecasting using different methods. We use the Japanese sub-national mortality data of $47$ prefectures from $1975$ to $2012$ as the training dataset and the data from the years $2013$ to $2022$ as testing data. We forecast the mortality rates for the testing data based on the training data using different methods. Then we calculate the annuity prices, $\text{PV}_{x,t}$ using the forecasts of mortality rates from different methods as well as the actual mortality rates. 

Table~\ref{tab:4} exhibits the average prices of annuities with annual payment $\$1$ and interest rate $2\%$ in Japan for some selected ages and years. Most forecasting methods tend to underestimate annuity prices, which is a common phenomenon in actuarial science, corresponding to the underestimated longevity risk \citep{ngai2011longevity}. Comparably, our methods slightly overestimate annuity prices. Further investigation of annuity prices reveals that the pricing errors of the proposed method are much lower than those of other methods. 

\begin{centering}
\renewcommand{\arraystretch}{0.68}
\setlength{\tabcolsep}{10pt}
\begin{longtable}{@{\extracolsep{0pt}}llrrrrrrr@{}}
\caption{Average annuity prices with annual payment $\$1$ and interest rate $2\%$ in Japan for selected ages and years. Annuity prices based on true mortality rates are labelled as "TRUE", annuity prices based on the proposed method are labelled as ``FDFM'', annuity prices based on high-dimensional functional time series are labelled as ``HDFTS'', annuity prices based on high-dimensional functional factor models are labelled as ``HDFFM'', annuity prices based on a matrix factor model are labelled as ``MFM'', annuity prices based on CPD tensor decompositions are labelled as ``CPD'' and annuity prices based on Tucker tensor decompositions are labelled as ``Tucker''.}\\
\toprule
\label{tab:4}
Sex & (Year, Age) & TRUE & FDFM & HDFTS & HDFFM & MFM & CPD & Tucker \\
\midrule
Female & (1980, 55) & $13.026$ & $13.032$ & $12.964$ & $12.940$ & $12.997$ & $13.040$ & $13.035$ \\ 
& (1985, 60) & $14.724$ & $14.731$ & $14.654$ & $14.627$ & $14.691$ & $14.740$ & $14.734$ \\ 
& (1990, 65) & $15.786$ & $15.794$ & $15.707$ & $15.676$ & $15.749$ & $15.805$ & $15.798$ \\ 
& (1995, 70) & $12.980$ & $12.989$ & $12.888$ & $12.852$ & $12.937$ & $13.001$ & $12.994$ \\ 
& (2000, 75) & $10.042$ & $10.053$ & $9.933$ & $9.890$ & $9.990$ & $10.067$ & $10.058$ \\ 
& (2005, 80) & $6.976$ & $6.990$ & $6.840$ & $6.787$ & $6.912$ & $7.007$ & $6.996$ \\ 
& (2010, 85) & $3.776$ & $3.795$ & $3.590$ & $3.518$ & $3.688$ & $3.819$ & $3.804$ \\ 
\midrule
Male & (1980, 55) & $10.081$ & $10.085$ & $10.057$ & $10.057$ & $10.067$ & $10.090$ & $10.090$ \\ 
& (1985, 60) & $11.698$ & $11.702$ & $11.670$ & $11.670$ & $11.682$ & $11.708$ & $11.708$ \\ 
& (1990, 65) & $12.863$ & $12.869$ & $12.830$ & $12.830$ & $12.844$ & $12.876$ & $12.875$ \\ 
& (1995, 70) & $10.411$ & $10.418$ & $10.371$ & $10.370$ & $10.388$ & $10.427$ & $10.426$ \\ 
& (2000, 75) & $7.981$ & $7.990$ & $7.928$ & $7.928$ & $7.950$ & $8.001$ & $8.001$ \\ 
& (2005, 80) & $5.571$ & $5.584$ & $5.495$ & $5.495$ & $5.527$ & $5.600$ & $5.599$ \\ 
& (2010, 85) & $3.161$ & $3.182$ & $3.031$ & $3.029$ & $3.085$ & $3.210$ & $3.209$ \\ 
\bottomrule
\end{longtable}
\end{centering}

To illustrate the financial impact of the mispricing on life insurers, consider the annuity pricing for females aged $75$ at year $2000$. The pricing error for the proposed method is \$0.011 per \$1 payment. However, the pricing errors for other competing methods range from \$0.016 to \$0.152 per \$1 payment. Suppose the annual payment for each individual is \$10,000, and $50,000$ people purchased this product. Then based on the proposed method, the capital that the insurer needs to reserve is reduced by at least $26$ million \Big($(\$0.016-\$0.011)\times 10,000\times50,000 = \$2.5 $million\Big), a significant saving to the life insurer.

\section{Conclusion}\label{sec:7}

We propose a factor model for HDFTS. By identifying homogeneity, we can reduce the HDFTS into an FTS of low dimensions. Then, a common functional front-loading can be determined to extract the temporal information in HDFTS into a low-dimensional FTS. The proposed model provides more flexibility, with certain choices of front and back-loadings; our model coincides with some of the existing factor models for HDFTS. 

Via a series of Monte-Carlo simulations, we demonstrate the performance of the proposed method for in-sample model fitting and out-of-sample forecast accuracy. An empirical study on the age-specific mortality rates of $47$ Japanese prefectures demonstrates the superiority of the proposed method over the existing models in forecasting. This result shows that the proposed model can extract temporal information more effectively.  

The proposed model can be viewed as a functional panel data model with interactive effects. We considered the multiplicity of time trend effects and population-specific effects instead of addition as in additive fixed effects models. 

In the current work, functional concurrent regression is performed for each $t \in\{1, \ldots, T\}$ to obtain the FTS of lower dimensions, which omits the temporal dependence of the functional observations. A possible future study could extend our functional concurrent regression to the FTS regression of \cite{pham2018methodology}. This awaits further investigation. 

\section*{Acknowledgment}

This research was financially supported by the Australian Research Council Discovery Project (DP230102250). The usual disclaimer applies.

\bibliographystyle{agsm}
\bibliography{FDFM}

\end{document}